\documentclass{aa}  
\usepackage{graphicx}
\usepackage{natbib}
\usepackage{multirow}
\usepackage{txfonts}

\usepackage[colorlinks, citecolor=blue, linkcolor=blue]{hyperref}
\hypersetup{colorlinks,breaklinks, linkcolor=blue,urlcolor=magenta, anchorcolor=blue,citecolor=blue}
\bibpunct{(}{)}{;}{a}{}{,} 

\newcommand{\be}{\begin{equation}}
\newcommand{\ee}{\end{equation}}

\def\gcm3{\hbox{g cm$^{-3}$}}       

\def\Mjup{\hbox{$\mathrm{M}_{\rm Jup}$}}

\def\Mearth{\hbox{$\mathrm{M}_{\oplus}$}}
\def\Rearth{\hbox{$\mathrm{R}_{\oplus}$}}

\newcommand{\logRHK}{$\log{R^{\prime}_{\rm HK}}$}

\newcommand{\serval}{\texttt{serval}}
\newcommand{\shaq}{\texttt{SHAQ}}
\newcommand{\sbart}{\texttt{S-BART}}

\newcommand{\gaia}{\textit{Gaia}}

\newcommand{\host}{KOBE-1}
\newcommand{\planetb}{KOBE-1\,b}
\newcommand{\planetc}{KOBE-1\,c}
\newcommand{\nspec}{99}
\newcommand{\perb}{8.5}
\newcommand{\perc}{29.7}
\newcommand{\timespan}{1180}
\newcommand{\nrv}{82}
\newcommand{\noutl}{17}
\newcommand{\mb}{8.80\,$\pm$\,0.76}
\newcommand{\mc}{12.4\,$\pm$\,1.1}

\begin{document}

\title{KOBE-1:
The first planetary system from the KOBE survey{
\thanks{Based on observations collected at the Centro Astron\'omico Hispano en Andaluc\'ia (CAHA) at Calar Alto (Spain), operated jointly by Junta de Andaluc\'ia and Consejo Superior de Investigaciones Cient\'ificas (IAA-CSIC).}}}
    \subtitle{Two planets likely residing in the sub-Neptune mass regime around a late K-dwarf}
    
   \author{
O.~Balsalobre-Ruza\inst{\ref{cab},\ref{ucm}},
J.~Lillo-Box\inst{\ref{cab}},
A.~M.~Silva\inst{\ref{caup},\ref{lisboa},\ref{porto}},
S.~Grouffal\inst{\ref{lam}},
J.~Aceituno\inst{\ref{caha},\ref{iaa}},
A.~Castro-González\inst{\ref{cab}, \ref{ucm}},
C.~Cifuentes\inst{\ref{cab}},
M.~R.~Standing\inst{\ref{esac}},
J.~P.~Faria\inst{\ref{geneve}},
P.~Figueira\inst{\ref{geneve}, \ref{porto}},
A.~Santerne\inst{\ref{lam}},
E.~Marfil\inst{\ref{upm}},
A.~Abreu\inst{\ref{atg}},
A.~Aguichine\inst{\ref{ucsc}},
L.~González-Ramírez\inst{\ref{cab}, \ref{ucm}},
J.~C.~Morales\inst{\ref{ice},\ref{ieec}},
N.~Santos\inst{\ref{porto}},
N.~Huélamo\inst{\ref{cab}},
E.~Delgado Mena\inst{\ref{cab}, \ref{caup}},
D.~Barrado\inst{\ref{cab}},
V.~Adibekyan\inst{\ref{caup}},
S.~C.~C.~Barros\inst{\ref{caup},\ref{porto}},
Á.~Berihuete\inst{\ref{cadiz}},
M.~Morales-Calderón\inst{\ref{cab}},
E.~Nagel\inst{\ref{inst:iag}},
E.~Solano\inst{\ref{cab}},
S.~G.~Sousa\inst{\ref{caup}},
J.~F.~Ag\"u\'i Fern\'andez\inst{\ref{caha}},
M.~Azzaro\inst{\ref{caha}},
G.~Bergond\inst{\ref{caha}},
S.~Cikota\inst{\ref{caha}},
A.~Fern\'andez-Mart\'in\inst{\ref{caha}},
J.~Flores\inst{\ref{caha}},
S.~G\'ongora\inst{\ref{caha}},
A.~Guijarro\inst{\ref{caha}},
I.~Hermelo\inst{\ref{caha}},
V.~Pinter\inst{\ref{caha}},
J.~I.~Vico~Linares\inst{\ref{caha}},
}

\titlerunning{\host: The first planetary system from the KOBE survey}
\authorrunning{O. Balsalobre-Ruza et al.}

\institute{
Centro de Astrobiolog\'ia (CAB), CSIC-INTA, ESAC campus, Camino Bajo del Castillo s/n, 28692, Villanueva de la Ca\~nada (Madrid), Spain \email{olga.balsalobre@cab.inta-csic.es}
\label{cab} 
\and
Departamento de F\'{i}sica de la Tierra y Astrof\'{i}sica, 
Facultad de Ciencias F\'{i}sicas, Universidad Complutense de Madrid, E-28040, Madrid, Spain
\label{ucm} 
\and
Instituto de Astrof\'{i}sica e Ci\^{e}ncias do Espa\c{c}o, Universidade do Porto, CAUP, Rua das Estrelas, 4150-762 Porto, Portugal\label{caup}
\and
Instituto de Astrof\'{i}sica e Ci\^{e}ncias do Espa\c{c}o, Universidade de Lisboa, Campo Grande, 1749-016 Lisboa\label{lisboa}
\and
Departamento de F\'{i}sica e Astronomia, Faculdade de Ci\^{e}ncias, Universidade do Porto, Rua do Campo Alegre, 4169-007 Porto, Portugal\label{porto}
\and
Aix Marseille Univ, CNRS, CNES, Institut Origines, LAM, Marseille, France\label{lam}
\and
Centro Astron\'omico Hispano en Andaluc\'\i a, Observatorio de Calar Alto, Sierra de los Filabres, 04550 G\'ergal, Almer\'\i a, Spain\label{caha}
\and
Instituto de Astrof\'isica de Andaluc\'ia, CSIC, Glorieta de la Astronom\'ia SN, 18008, Granada, Spain\label{iaa}
\and
European Space Agency (ESA), European Space Astronomy Centre (ESAC), Camino Bajo del Castillo s/n, E-28692 Villanueva de la Cañada, Madrid, Spain\label{esac}
\and
Observatoire Astronomique de l’Universit\'{e} de Gen\`{e}ve, Chemin Pegasi 51b, 1290 Versoix, Switzerland\label{geneve}
\and
Departamento de Ingenier\'{i}a Topogr\'{a}fica y Cartograf\'{i}a, E.T.S.I. en Topograf\'{i}a, Geodesia y Cartograf\'{i}a, Universidad Polit\'{e}cnica de Madrid, 28031 Madrid, Spain\label{upm}
\and
ATG Europe for European Space Agency (ESA), Camino bajo del Castillo, s/n, Urbanizacion Villafranca del Castillo, Villanueva de la Cañada, 28692 Madrid, Spain\label{atg}
\and
Department of Astronomy and Astrophysics, University of California, Santa Cruz, CA 95064, USA\label{ucsc}
\and
Institut de Ci\`encies de l'Espai (ICE, CSIC), Campus UAB, c/ Can Magrans s/n, E-08193 Bellaterra (Barcelona), Spain\label{ice}
\and
Institut d'Estudis Espacials de Catalunya (IEEC), C/Esteve Terradas, 1, Edifici RDIT, Campus PMT-UPC, E-08860 Castelldefels (Barcelona), Spain\label{ieec}
\and
Depto. Estadística e Investigación Operativa, Universidad de Cádiz, Avda. República Saharaui s/n, 11510 Puerto Real, Cádiz, Spain\label{cadiz}
\and
Universit\"at G\"ottingen, Institut f\"ur Astrophysik und Geophysik, Friedrich-Hund-Platz 1, 37077 G\"ottingen, Germany\label{inst:iag}
}

   \date{}

  \abstract
   {K-dwarf stars are promising targets in the exploration of potentially habitable planets. Their properties, falling between G and M dwarfs, provide an optimal trade-off between the prospect of habitability and ease of detection. The KOBE experiment is a blind-search survey exploiting this niche, monitoring the radial velocity of 50 late-type K-dwarf stars. It employs the CARMENES spectrograph, with an observational strategy designed to detect planets in the habitable zone of their system.}
   {In this work, we exploit the KOBE data set to characterize planetary signals in the K7\,V star \object{HIP~5957} (\host) and to constrain the planetary population within its habitable zone.} 
   {We used \nrv\ CARMENES spectra over a time span of three years. We employed a generalized Lomb-Scargle periodogram to search for significant periodic signals that would be compatible with Keplerian motion on KOBE-1. We carried out a model comparison within a Bayesian framework to ensure the significance of the planetary model over alternative configurations of lower complexity. We also inspected two available TESS sectors in search of planetary signals.}
   {We identified two signals: at $P_{\rm b}$ = \perb\ d and $P_{\rm c}$ = \perc\ d. We confirmed their planetary nature through ruling out other non-planetary configurations.
   Their minimum masses are \mb\,M$_{\oplus}$ (\planetb{}), and \mc\,M$_{\oplus}$ (\planetc{}), corresponding to absolute masses within the planetary regime at a high certainty ($>99.7$\%). By analyzing the sensitivity of the CARMENES time series to additional signals, we discarded planets above 8.5\,M$_{\oplus}$ within the habitable zone. We identified a single transit-like feature in TESS, whose origin is still uncertain, but still compatible within 1$\sigma$ with a transit from planet c.}
   {The \host\ multi-planetary system, consisting of a relatively quiet K7-dwarf hosting two sub-Neptune-minimum-mass planets, establishes the first discovery from the KOBE experiment. We have explored future prospects for characterizing this system, concluding that \gaia{} DR4 will be insensitive to their astrometric signature. Meanwhile, nulling interferometry with the Large Interferometer For Exoplanets (LIFE) mission could be capable of directly imaging both planets and characterizing their atmospheres in future studies.}
   

   \maketitle
%
\section{Introduction}
\label{sec:intro}

The detection of new extrasolar planets has been led by two main techniques to date, transit photomery and radial velocity (RV). While the transit method has been very efficient from space-based missions, such as Kepler \citep{borucki10} and TESS \citep{ricker14}, thanks to its brute-force approach to observing millions of stars, the RV searches allow for a more focused and flexible search. In contrast to transits, the nature of the RV method allows for the detection of planets with a relatively high orbital inclination with respect to the line of sight and  at longer orbital periods as well. However, one of the main handicaps of this technique is the large amount of observing time required in ground-based telescopes to reach a final confirmation. Nonetheless, several surveys have been granted time in recent decades to apply high-precision and stabilized spectrographs to  these blind-search surveys in specific niches. Some examples are the High Accuracy Radial velocity Planet Searche (HARPS) search for southern extrasolar planets (e.g., \citealt{bonfils05, pepe11, udry19}),  GAPS collaboration with HARPS-N in the northern hemisphere (\citealt{poretti16}), Calar Alto high-Resolution search for M-dwarfs with Exoearths with Near-infrared and optical \'Echelle Spectrographs (CARMENES) survey searching for planets around mid- to late-M-dwarfs (\citealt{quirrenbach14,ribas23}), and Echelle SPectrograph for Rocky Exoplanets and Stable Spectroscopic Observations (ESPRESSO) guaranteed time observations (e.g., \citealt{pepe21}, \citealt{lillo-box21c}).

One particular parameter space that is especially better covered by the RV technique than by transits is the regime corresponding to warm stellar insolations. Within this domain, there is the so-called habitable zone (hereafter, HZ), defined by \cite{kasting93} and \cite{kopparapu13} as the region around a star where a rocky planet with certain atmospheric properties could sustain liquid water on its surface. This concept has been largely used as the roadmap\footnote{See, for instance, the \textit{Follow the water} premise of NASA (\citealt{briggs00,hubbard02}).} for exoplanet missions (e.g., Kepler or PLATO). The transit probability of a planet in such regime (especially for FGK stars) is certainly low (e.g., 0.5\% and 1.2\% for a planet in the inner and outer edge, respectively, of the HZ of a K7 dwarf); meanwhile,  RV surveys can be specifically designed to increase the efficiency of exoplanet detection in this regime. 

This is indeed the case of the K-dwarfs Orbited By habitable Exoplanets (KOBE) experiment\footnote{\url{https://kobe.caha.es}} \citep{lillo-box22}, which is a legacy program of the Calar Alto observatory (Spain) using the CARMENES instrument at the 3.5m telescope. Its main goal is to look for temperate planets around late K-dwarfs, since these stars provide the ideal conditions for habitable planets from both the detection and the astrobiological point of view (e.g., \citealt{isaacson10, cuntz16, arney19, lillo-box22}). Planets within the HZ of these stars will be amenable for atmospheric characterization even if they do not transit thanks to future missions such as the proposed LIFE (Large Interferometer For Exoplanets) project \citep{quanz21}.

According to the occurrence rate study by \cite{kunimoto20} based on Kepler data, K-dwarf stars are statistically expected to harbor around 1.84 planets with orbital periods under 200 d. Among them, around one-third (0.56 planets per star) are expected to reside within the HZ regime. Consequently, although the KOBE experiment is designed to detect planets in this temperate region, others are also expected to be found. A particularly interesting parameter space that KOBE observations can explore is the sub-Neptune regime. Detecting and characterizing these planets is currently one of the endeavors in the field since they mark the transition between gaseous and rocky planets. The processes driving the transition between those classes or contributing to their formation and evolution, remain under debate. Therefore, discovering new and uncommon sub-Neptune planets in sparse regions of the parameter space such as within the hot Neptune desert (e.g., \citealt{hacker24}), within the radius valley (e.g., \citealt{osborne24}), or low-density super-Earths (e.g., \citealt{castro-gonzalez23}) is of significant value. 

In this paper, we present the detection and subsequent confirmation of two planetary signals with minimum masses in the sub-Neptune regime orbiting the star HIP\,5957 (hereafter \host).
In Sect.~\ref{sec:obs}, we present the observations that led to the discovery of the signals. In Sect.~\ref{sec:stellarprop} we study the properties of the stellar host, including its activity imprints. Section~\ref{sec:analysis} presents the analysis of the RV data (Sect.~\ref{sec:RVanalysis}) and a search for transiting counterparts in TESS (Sect.~\ref{sec:TESSanalysis}). We then discuss the results in Sect.~\ref{sec:discussion}, including a confirmation of the planets  (Sect.~\ref{sec:verification}) following the Exoplanet Confirmation Protocol premisses (Lillo-Box et al., in prep.), along with an analysis of the two planetary signals in the context of the exoplanet population (Sect.~\ref{sec:warsemnep}), a study of the sensitivity limits focused on the HZ of the star (Sect.~\ref{sec:sensitivity}), and a discussion of the prospects for further characterization (Sect.~\ref{sec:prospects}). We  conclude with some remarks in Sect.~\ref{sec:conclusions}.

\section{Observations}
\label{sec:obs}

\subsection{CARMENES high-resolution spectroscopy}
\label{sec:carm}
We used the CARMENES (\citealt{quirrenbach14}) instrument to monitor the RV of \host\ (\object{GJ~9048}, \object{TIC~16917838}, \object{HIP~5957}). These observations are part of the KOBE experiment survey (PI: Lillo-Box; \citealt{lillo-box22}), running at the Calar Alto Observatory since January 2021. The CARMENES instrument is fed by the 3.5-meter telescope of this observatory and is located in an isolated chamber with active environmental control. The spectrograph is housed inside a vacuum vessel stabilized in temperature down to the 10\,mK level in 24\,hours \citep{quirrenbach14}. The light collected by the telescope through fiber A is split into the optical (hereafter, VIS) and the near-infrared (NIR) channels, both located in separate chambers and vacuum vessels. In order to monitor intra-night changes in the wavelength solution, we used a Fabry-Perot (FP) injected into fiber B and projected into the inter-order regions of fiber A in the VIS and NIR detectors.  

We monitored \host\ for over three years (\timespan\ d in total), with an average cadence per observing season of one spectrum every five days. We obtained a total of \nspec\ spectra within this period. The exposure times typically varied from 500\,s to 1800\,s depending on weather conditions, reaching an average signal to noise natio (S/N) per pixel at 650\,nm of 93. The basic data reduction was performed at the observatory by the CARACAL pipeline (\citealt{zechmeister14, bauer15}), which performs the basic bias, flat fielding, background corrections, and extraction of the spectra from the individual orders. The pipeline also estimates the RV drift from the FP orders, and the barycentric Julian date (BJD) of the observation. We calculated the Barycentric Earth RV (BERV) correction to apply it to the wavelength solution.

The extracted VIS spectra were then used to measure the stellar RV. In a first step, we used our own developed pipeline \shaq\ \citep{lillo-box22} based on the cross-correlation technique \citep{baranne96}. \shaq\ uses the publicly available binary masks from the ESPRESSO instrument pipeline (\citealt{pepe21}). In this case, we used the M0 binary mask weighted by the line depth and a selection of the CARMENES orders, removing those heavily affected by telluric bands or showing ${\rm S/N} < 20$. From the cross-correlation function (CCF) of each spectrum, we extracted the RV as the mean of the Gaussian profile, its full-width at half maximum (FWHM), and its contrast and the bisector-span (BIS),  computed as detailed in \cite{lafarga20}.

The RVs extracted from \shaq\ were used as prior information in our main RV extraction pipeline optimized for high-resolution spectroscopy, \sbart\footnote{Available at \url{https://github.com/iastro-pt/sBART}} \citep{silva22}. \sbart\ uses the template matching approach in a Bayesian framework, providing a straightforward and consistent method to characterize the RV posterior probability associated with each observation. Furthermore, the code assumes a common RV shift across all wavelengths, reflecting the RV signal injected by planetary companions. Prior to the RV extraction, a pre-processing stage was applied to each spectrum. For that end, the quality control flags were evaluated and any non-physical fluxes were rejected (e.g., null values, wavelength ranges affected by chromospheric activity; see \citet{silva22} for further details). Telluric features were then handled by masking out the spectral regions where they are present. This was accomplished through the construction of a binary mask from a synthetic transmittance profile of Earth's atmosphere using \texttt{Telfit}\footnote{Available at \url{https://github.com/kgullikson88/Telluric-Fitter}} (\citealt{2014ascl.soft05002G}). To ensure that the extracted RVs were informed by the same spectral information, \sbart\ only used the wavelength regions that are common to all observations in the dataset. Finally, \shaq\ RVs were used for the construction of a preliminary stellar template, which was then used to extract \sbart\ RVs. Then, a second stellar template was constructed from the newly derived \sbart\ RVs and used to extract the final set of RVs.

To compute additional activity indicators and as cross-check, we also used the \serval\ pipeline\footnote{Available at \url{www.github.com/mzechmeister/serval}} \citep{zechmeister18}. This code uses the classical template matching approach and provides RVs from both the VIS and the NIR channels. This pipeline also provides relevant activity indicators that had been demonstrated to be useful to pinpoint the stellar noise due to magnetic and granulation effects. These indicators include the chromatic index (CRX), the differential line width (dLW), and several line indices (i.e., integrated fluxes), such as H$\alpha$, the calcium infrared triplet (CaIRT$_1$ and CaIRT$_2$), and the sodium doublet (NaD$_1$ and NaD$_2$). The time series for the \serval\ and the CCF spectroscopic indicators are shown in Table~\ref{tab:RVindicators}.

Despite its thermal and environmental isolation, CARMENES is known to suffer from night-to-night RV jumps at the level of few m/s. Since the source of these offsets is not understood yet (\citealt{ribas23}), the determination of the so-called nightly zero points (hereafter NZPs) is a critical step towards achieving the necessary accuracy to detect extrasolar planets. The NZPs were then calculated every night following the same procedure explained in \cite{trifonov18}. To that end, we monitored at least two standard stars (typically three) every night that a KOBE target is observed with CARMENES, with a total of six standards along the year. We also included KOBE targets (i.e., non-standard stars) in the NZP computation, provided that at least five were observed that night and their RV root mean square (RMS) were below 10\,m\,s$^{-1}$ to balance the potential presence of additional signals. We note that we left out 
the particular KOBE target for which the NZP correction were to be applied (in this case, \host). The NZPs obtained with \sbart\ have 51\% smaller RMS than that from \serval, and it is $\sim$37\% of that from \shaq\ (for details, see Appendix\,\ref{sec:pipes}). For these reason, along with the proven improved precision of \sbart\ as compared with both classic template-matching and CCF-based algorithms in several publications (e.g., \citealt{faria22, palethorpe24, passegger24, suarezmascareno24}), we only used the RV time series and NZP corrections extracted with \sbart\ in this work.

We discarded \noutl\ spectra based on different reasons: 
(i) four of them were flagged as having an overly large drift correction; this is the result of CARACAL computing the value from calibration files of past nights when the corresponding one is incomplete; 
(ii) two had a S/N at 7500\,\r{A} below 20, which translates into an RV uncertainty ($\sigma_{\rm RV}$) larger than 10\,m\,s$^{-1}$; 
(iii) two nights lacked a proper NZP correction available since no standards were observed due to adverse weather conditions;
and (iv) nine measurements were ruled out to ensure no contamination due to the Moon illumination. These nights  simultaneously satisfy that the angular separation between \host\ and the Moon was below $80^{\circ}$, the Moon illumination was above 40\%, and the difference between the BERV and the \host\ velocities was below three times the FWHM of our typical line width (i.e., below 21 km\,s$^{-1}$). In Appendix~\ref{sec:moon}, we demonstrate that these potentially Moon contaminated data do not show an effect in the RV signals found. However, we did not include them in the analyzed time series to maintain a conservative overall stance. 
The final RV dataset, composed of \nrv\ measurements extracted with \sbart, is provided in Table~\ref{tab:rv_sbart}, together with the associated drift and NZP values. 
In Table~\ref{tab:F3}. we identify the dates discarded from the dataset. 

The final \sbart\ RV time series and all computed activity indicators together with their RV correlation and generalized Lomb-Scargle (GLS, \citealt{zechmeister09}) periodograms are shown in Fig.~\ref{fig:actind}. The colored (and grey) lines in the periodograms correspond to the detrended (and non-detrended) time series using a second-order polynomial (see grey line in the left panel), whose nature is discussed in Sect.~\ref{sec:RVanalysis}. 
The RVs show two peaks that stand above the false alarm probability (FAP) of 1\%, at \perb\ and \perc\,d, when subtracting the long-term trend (blue and green lines, respectively). The same periodicities were also found in the \serval\ RV time series as shown in Appendix~\ref{sec:pipes}. Aliases from the sampling were ruled out as their origin since they do not appear in the spectral window function (see Fig.~\ref{fig:E2}).
On that side, activity indicators do not show any significant power at those frequencies and there is no correlation with the RVs, thereby suggesting that these signals might be produced by two planetary-mass companions to \host. Some caution must be taken as the outer periodicity (\perc~d) is close to the synodic period of the Moon (i.e., 29.53\,d). As detailed above and further explored in Appendix~\ref{sec:moon}, the dataset is clean from any potential Moon contamination and we confidently ruled out lunar contamination as the cause for this periodic signal.

\subsection{HARPS-N}
\label{sec:obs_hn}
We obtained one spectrum of \host{} using the HARPS-N instrument \citep{cosentino14} installed at the 3.6\,m Telescopio Nazionale di Galileo (TNG) located in the Observatorio del Roque de los Muchachos at La Palma (Spain). HARPS-N is a temperature- and pressure-stabilized high-resolution spectrograph with a resolution of 115\,000 and a wavelength coverage between 385-691\,nm. We obtained this single spectrum on the night of 2-June-2024 under good weather conditions and an airmass of 1.68. We used an exposure time of 1800\,s and obtained a S/N spectrum of 86 at 650\,nm. This spectrum was processed by using the online pipeline at the observatory, which performs the basic reduction and spectrum extraction. We used this spectrum to derive the stellar rotational period from the chromospheric activity index \logRHK{} (\citealt{noyes84}), detailed\ in Sect.~\ref{sec:act}.

\subsection{TESS photometric time series}
\label{sec:tess-data}

\host\ (\gaia{}\,DR3\,294517800251711616, TIC\,16917838) was observed by the Transiting Exoplanet Satellite Survey (TESS) space-based mission \citep{ricker14} in sectors 17 and 57 with 2-minute cadence using camera \#1 in both cases. We retrieved the data from the MAST archive\footnote{\url{https://archive.stsci.edu}} through the \texttt{lightkurve}\footnote{Available at \url{https://docs.lightkurve.org/index.html}} package \citep{lightkurve}. The data were automatically processed by the official Science Processing Operations Center (SPOC) pipeline \citep{jenkins16}, producing instrumentally corrected light curves in the so-called Pre-Data Conditioned Search Aperture Photometry (PDCSAP) format. 

We checked for contaminants and close companions to \host\ by using the \texttt{tpfplotter}\footnote{Available at \url{https://github.com/jlillo/tpfplotter}} \citep{aller20} and \texttt{tess-cont}\footnote{Available at \url{https://github.com/castro-gzlz/tess-cont}} (\citealt{castrogonzalez24}) algorithms. The two panels in Fig.~\href{https://zenodo.org/records/14511891}{E.3} show the average stamps of sectors 17 and 57, including the location of the target, the aperture used by the SPOC pipeline to extract the light curve, and the \gaia{} DR3 sources in the field up to a contrast magnitude of $\Delta G<8$~mag. As shown in Fig.~\ref{fig:E3}, there is only one nearby source (labeled as \#2 and corresponding to \gaia{} DR3 294517868971187456 or TIC 16917841) observed in the stamps, which lies within the SPOC aperture with a projected separation of 58$\arcsec$ and a magnitude contrast of $\Delta G=5.3$~mag. We found that the contribution from this source to the total aperture flux is negligible (0.24$\%$), which is expected due to the large contrast and its location near the edge of the aperture as seen in other studies \citep[e.g.,][]{2020A&A...642A.121L,2021A&A...654A..60L,2022MNRAS.509.1075C}.

It is still important to study bright sources farther away than the TESS stamps since they have been found to affect the light curves of several targets (e.g., \citealt{aller24}). In the case of \host, there is a bright companion object ($\Delta G=1.1$~mag, \gaia{}\,DR3\,294514982753165696 or TIC\,16917834) located at 162$\arcsec$ to the south-east, which we refer to as Star\#3. We found that the potential contamination caused by this star to the SPOC aperture is 0.04$\%$. As we show in Fig.~\ref{fig:E4}, the total contaminant flux only corresponds to a 0.3$\%$ of the flux falling inside the aperture, which indicates that the photometric features of the nearby stars surrounding \host\ do not affect its photometry. That negligible contaminant flux mainly comes from Star\#2 (82\%) and \#3 (12\%) fluxes, while the remaining 6$\%$ comes from smaller contributions of 68 faint nearby stars.

\subsection{ASAS-SN long-term photometry}

\host{} has been observed by seven cameras of the All-Sky Survey for Supernovae \citep[ASAS-SN,][]{2014ApJ...788...48S} since 2012 using telescopes from different latitudes.
In Table~\ref{tab:F4}, we summarize the details of these observations.

We extracted the photometry of \host{} through the Sky Patrol web interface\footnote{\url{https://asas-sn.osu.edu/}} by selecting the multilevel perception (MLP) extraction (\citealt{winecki24}), which generates unbiased photometry for stars ranging  $g \simeq$ 4 - 14 mag, and also provides better results than the standard ASAS-SN pipelines as tested by the authors. We followed the approach described in \citet{castro-gonzalez23} to perform periodic coordinate corrections to minimize potential flux loses due to the proper motion of our target star. 
We discarded epochs with flux values below the estimated 5$\sigma$ detection limit for the target location.

In Fig.~\ref{fig:E5}, we show the complete ASAS-SN photometric time series (except for the camera \texttt{bf}, which only acquired six data points), with their \texttt{GLS} periodograms, and the data folded in phase to the corresponding maximum power periods. We identified a prominent signal detected in four independent cameras at $\sim$29.6\,d 
(coinciding with the lunar synodic month). Since we also found signals at 27.3\,d (corresponding to the lunar sidereal month) in three of these cameras, we interpreted that the monthly luminosity variations of the sky caused by the Moon were affecting these photometric time series as found in other works (e.g., Benatti et al., in prep.).  
We did not find any additional significant signals in those four periodograms; whereas cameras \texttt{bj} and \texttt{br} show maximum power periods at 77 and 3\,d, with no significant FAPs (0.5 and 2$\%$, respectively). We remark here that the CARMENES spectra potentially contaminated by the Moon illumination were ruled out (see Sect.~\ref{sec:carm}) and we demonstrated that there is no equivalent effect in our RV time series (see Appendix~\ref{sec:moon}).

\section{Stellar properties}
\label{sec:stellarprop}

\subsection{Physical properties}

\host\ is a relatively bright ($V \simeq$ 10\,mag), nearby ($23.880 \pm 0.012$\,pc), high proper motion ($\mu \simeq$ 440\,mas\,yr$^{-1}$) late-K to early-M dwarf.
It does not have known close ($a \lesssim$ 200\,au or $\rho \lesssim$ 8.4\arcsec) stellar companions and none of the statistical metrics from \gaia{} DR3, as presented in \cite{cifuentes24}, would suggest an unresolved multiplicity.
We determined the components of the Galactocentric space velocity, $UVW$, using the \textsc{SteParKin} code\footnote{Available at \url{https://github.com/dmontesg/SteParKin}.} \citep{montes01}.
With these parameters, the systemic RV ($\gamma$), and the parallax ($\varpi$), we used the same tool to assign the star to the kinematic population of the Galactic thin disc.

The co-adding of the almost three-year CARMENES spectra from \host\ allowed us to derive its stellar atmospheric parameters: effective temperature ($T_{\rm eff}$), surface gravity ($\log{g}$), and iron abundance ([Fe/H] as a proxy for metallicity).
We followed the same approach as in \cite{lillo-box22}, which is based on the one used for the CARMENES GTO sample (\citealt{Marfil2021}). 
In short, we performed the analysis with a grid of BT-Settl model atmospheres (\citealt{allard12}) and the \textsc{SteParSyn} code (\citealt{Tabernero2022}), a Bayesian implementation of the spectral synthesis technique. 
We safely assumed a fixed $\varv\sin{i}=2~\rm km\,s^{-1}$ as an upper limit since no significant rotation was detected in the stellar spectrum \citep[see][]{reiners18}.
Our value for $T_{\rm eff}$ (4135\,$\pm$\,36\,K) is compatible with previous determinations \citep[e.g., 4157\,$\pm$\,98\,K from][]{gaidos14}.

We derived the bolometric luminosity ($L_{\star}$) by fitting photometric data of \host\ in several passbands (from SLOAN/SDSS $u'$ to AllWISE $W4$) to BT-Settl (CIFIST) synthetic models \citep{Baraffe98} with solar metallicity ([Fe/H] = 0.0) using Virtual Observatory Spectral energy distribution Analyzer \citep[{VOSA};][]{bayo08}.
From $L_{\star}$, we calculated the stellar radius using the Stefan-Boltzmann law. By using the mass-radius relation from \cite{Schweitzer19} (see their Eq.~6), we computed the stellar mass. 
This relation is based on studies of detached, double-lined, double-eclipsing, main sequence M-dwarf binaries from the literature and is applicable across a broad range of metallicities for stars older than several hundred million years.
The fundamental parameters obtained are compatible with a K7\,V star \citep{cifuentes20}, classification that we support spectroscopically in this work.
These stellar properties are summarized in Table~\ref{tab:stellar_param}.

\subsection{Stellar activity and rotation}
\label{sec:act}
We used the HARPS-N spectrum presented in Sect.~\ref{sec:obs_hn} to compute the \logRHK{} from calcium lines. We computed the Ca II H\&K index (named I\_caII in \texttt{ACTIN2}) by using the \texttt{ACTIN2} tool \citep{actin}, obtaining a value of $0.865 \pm 0.006$. We then used the \texttt{pyrhk} wrapper to determine the \logRHK{} from this index. 
Since the B-V color of this target is 1.379, we applied the \cite{rutten84} bolometric calibrations for main-sequence stars to use the \cite{noyes84} procedure. By doing so, we obtained a \logRHK$=-4.896 \pm 0.003$ dex, indicating a potentially low activity level from this star \citep[e.g., ][]{brown22}. The inferred rotation period of the star is $P_{\rm rot} = 37.4 \pm 6.0$\,d, obtained from the empirical relations presented in \cite{suarez-mascareno16} (assuming the corresponding K-dwarf coefficients based on the determined effective temperature for \host{}). This result is reinforced by the GLS analysis presented in Fig.~\ref{fig:actind}, where the 37.4\,d periodicity and its first harmonic ($P_{\rm rot}/2 \sim 18.7$\,d) appear in the time series of some of the \serval\ activity indicators (such as the dLW and H$\alpha$, but even more significantly in CaIRT$_{\rm 1}$ and CaIRT$_{\rm 2}$). 

\section{Analysis and results}
\label{sec:analysis}

\subsection{Radial velocity}
\label{sec:RVanalysis}

We modeled the \sbart\ RV measurements using $N_p$ Keplerians.
There is a quadratic long-term trend that we interpret as either the influence of a long-period companion for which there is no additional evidence (Sect.~\ref{sec:stellarprop}), systematic effect, or magnetic cycle that would be temporally unresolved (with its period, $P_{\rm cyc}$, being at least twice the time span of our observations). The two latter interpretations are reinforced by the observation of other long-trends in some activity indicators, including FWHM, dLW, CaIRT$_{\rm 1}$, and CaIRT$_{\rm 2}$ (see Fig.~\ref{fig:actind}).

Each Keplerian models one planetary signal and is described by its five standard parameters: RV semi-amplitude, $K$; orbital periodicity, $P$; conjuction time, $T_0$; eccentricity, $e$; and argument of periastron, $\omega$. Four additional parameters account for the second-order polynomial (the slope $m$, and the quadratic term $q$), the systemic velocity ($\gamma$), and an extra RV noise ($\sigma_{\rm jit}$) that is added quadratically to $\sigma_{\rm RV}$. Therefore, the model involves a total of 5\,$N_p$ + 4 free parameters. In the following subsections, we show the analysis of the CARMENES data set through different Bayesian approaches.

\subsubsection{Diffusive nested sampling}
\label{sec:kima}

We ran the \texttt{kima}\footnote{Available at \url{https://www.kima.science/docs/}} package (\citealt{2018JOSS....3..487F}) to explore how many signals are justified by the data ($N_p$), since this code uses the number of Keplerians as a free parameter instead of fixing it. We defined the likelihood distribution as a Student-t, and set uninformative prior distributions as detailed in Table~\ref{tab:F6}, allowing for up to three Keplerians ($N_p < 3$). The \texttt{kima} code uses the diffusive nested sampling algorithm (\citealt{2010ascl.soft10029B}) to infer the posterior distributions of the model parameters. Additionally, it provides the Bayesian evidence ($\mathcal{Z}$) to compare the competing models (i.e., $N_{p+1}$ against $N_{p}$).

From sampling with 150\,000 steps, we found a significant detection of two Keplerians of similar amplitudes ($K \sim$\,3.3\,m\,s$^{-1}$), with orbital periods matching those identified in the GLS periodogram ($P_{\rm b}$\,=\,\perb\,d and $P_{\rm c}$\,=\,\perc, see Fig.~\ref{fig:actind}), and eccentricities compatible with zero. The evidence for this two-planets model (hereafter, we refer the models as ``$N_p$p'' so ``2p'' corresponds to a two-planets model) is $\ln{\mathcal{Z_{\rm 2p}}} = -242.1$, which is strongly preferred over the null hypothesis (no planets) with a logarithm of the Bayes Factor (defined as $\Delta\ln{\mathcal{Z_{\rm ip,\,jp}}} = \ln{\mathcal{Z_{\rm ip}}} - \ln{\mathcal{Z_{\rm jp}}}$) of $\Delta\ln{\mathcal{Z_{\rm 2p,\,0p}}} = 10.9$. Here,   $\Delta\ln{\mathcal{Z}}$~$>$~5 is the typical threshold adopted by the community to consider a planetary detection (e.g., \citealt{faria22}, \citealt{beard22}) since it is recognized as a very strong evidence (e.g., \citealt{kass95}). There is also a decisive evidence for the existence of two planets as compared with a single one with $\Delta\ln{\mathcal{Z_{\rm 2p,\,1p}}} = 9.6$.

\subsubsection{MCMC and importance sampling}
\label{sec:emcee}

In this case, we employed the \texttt{radvel}\footnote{Available at \url{https://radvel.readthedocs.io/en/latest/}} python package (\citealt{2018PASP..130d4504F}) to model the Keplerians. We opted for a parametrization with $\sqrt{e}\cos{\omega}$ and $\sqrt{e}\sin{\omega}$ (instead of $e$ and $\omega$ separately) as suggested by \cite{eastman13} to avoid a truncated marginalized posterior distribution of $e$ at 0.

The prior distributions were chosen to be data-driven. In particular, we used a modified log-uniform distribution for the $\sigma_{\rm jit}$ with knee at the mean $\sigma_{\rm RV}$ ($\bar{\sigma}_{\rm RV}$ = 1.9\,m\,s$^{-1}$), and allowing for values up to the range of the data ($\Delta$RV\,=\,RV$_{\rm max}$\,--\,RV$_{\rm min}$\,=\,24.9\,m\,s$^{-1}$). 
For $P,$ we selected a log-uniform distribution from 1.1\,d (above one day to avoid fake signals caused by the sampling) to 100\,d since no (even marginal) signals were found at higher periodicities according with the GLS periodogram (see Fig.~\ref{fig:actind}). 
We used uniform distributions for the rest of parameters:
the slope prior ranged from \mbox{--0.05} to 0.05 m\,s$^{-1}$\,d$^{-1}$, with the quadratic trend from \mbox{--0.01} to 0.01 m\,s$^{-1}$\,d$^{-2}$; 
both the eccentricity parameter priors ($\sqrt{e}\cos{\omega}$ and $\sqrt{e}\sin{\omega}$)  were valid for for their whole domain, $\mathcal{U}(-1,\,1)$; the 
$\gamma$ prior ranged from its most extreme values (RV$_{\rm min}$\,=\,$-$23\,306\,m\,s$^{-1}$ and RV$_{\rm max}$\,=\,$-$23\,281\,m\,s$^{-1}$), taking into account $\Delta$RV;  thereby giving $\mathcal{U}$(RV$_{\rm min} - \Delta$RV, RV$_{\rm max} + \Delta$RV). The 
$K$ prior distribution went from 0 to 14\,m\,s$^{-1}$, which is more than three times the RV RMS, while the $T_{0}$ prior covered 100\,d (maximum period allowed) from the first observing date.

We sampled the posterior distributions through the Markov chain Monte Carlo (MCMC) affine invariant ensemble sampler \texttt{emcee}\footnote{Available at \url{https://emcee.readthedocs.io/en/stable/}} (\citealt{foreman-mackey13}). We used four times the number of parameters  as the number of walkers and we ran 140\,000 steps as a burn-in phase. Subsequently, we run a second phase with half the steps of the initial one (70\,000) in a narrower region of the parameter space around the maximum a posteriori solution from the first phase (as suggested in the \texttt{emcee} documentation\footnote{\url{https://emcee.readthedocs.io/en/stable/tutorials/line/}}). We checked the convergence so the chain lengths surpassed 50 times the autocorrelation time. We used the chain from the second phase to get the marginalized posterior distributions and to compute the model evidence, as detailed below.

\begin{figure}[]
\centering\includegraphics[width=0.5\textwidth{}]{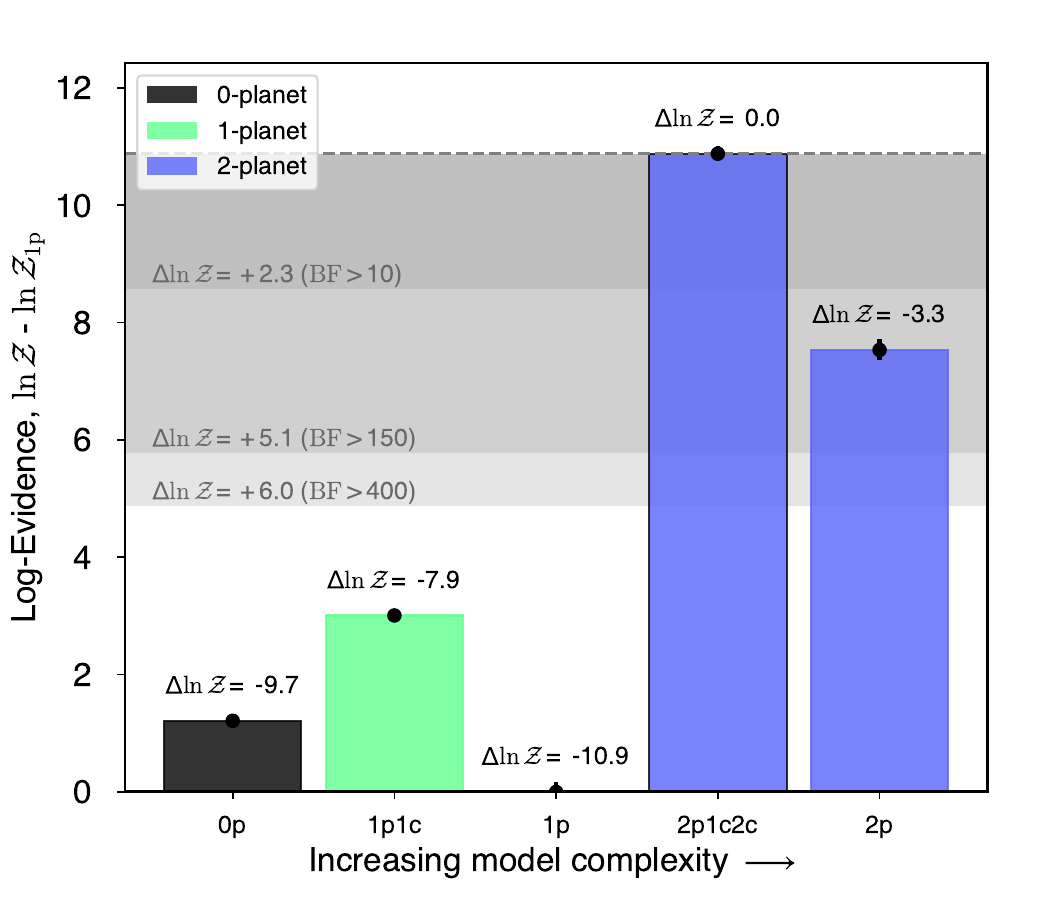}
\caption{Comparison of the logarithm of the Bayesian evidence ($\ln{\mathcal{Z}}$) for the tested RV models. The evidence difference ($\Delta\ln{\mathcal{Z}}$) with the best model, or Bayes Factor (BF), is shown at the top of the bar for each model.
In grey are displayed different $\Delta\ln{Z}$ from the 2p1c2c model marking the strong evidence against competing models.}
\label{fig:BF}
\end{figure}

We explored five models by varying the number of Keplerians ($N_p$ from 0 to 2) and fixing (or not) their architectures to circular orbits. We now use the labels ``$N_p$p[$\mathcal{P}_i$c]'', where $\mathcal{P}_i$ are the identifiers of the planets with assumed circular orbits (e.g., ``2p1c'' corresponds to a two-planet model with only the first planet in fixed circular orbit). To perform the same model comparison as in Sect.~\ref{sec:kima}, we computed the Bayesian log-evidence using the \texttt{bayev}\footnote{Available at \url{https://github.com/exord/bayev/tree/master}} (\citealt{2016A&A...585A.134D}) package based on the importance sampling estimator introduced by \cite{perrakis14}, feeding this code with the chain from the second phase of the MCMC. In Fig.~\ref{fig:BF}, we show the logarithm of the Bayesian evidence ($\ln\mathcal{Z}$) for the competing models. As in Sect.~\ref{sec:kima}, we found that the preferred model is 2p1c2c, with all orbital parameters compatible with those previously found in the \texttt{kima} analysis. This model has an absolute evidence of $\ln{\mathcal{Z_{\rm 2p1c2c}}}$ = --250.0, and a logarithm of the likelihood $\ln{\mathcal{L_{\rm 2p1c2c}}}$ = --508.4. We found log-Bayes factors of $\Delta\ln{\mathcal{Z}_{\rm 2p1c2c,\,0p}}$ = 9.7, and $\Delta\ln{\mathcal{Z}}_{\rm 2p1c2c,\,1p1c}$ = 7.9. Eccentric orbits were not preferred over circular ones, with $\Delta\ln{\mathcal{Z}}_{\rm 2p,\,2p1c2c}$ = --3.3; thus, adding two additional parameters was not justified. Both free eccentricity models, 1p and 2p, converged to wide posteriors for both $\sqrt{e}\cos\omega$ and $\sqrt{e}\cos\omega$ parameters, but are compatible with circular orbits.

\begin{figure*}[]
\centering\includegraphics[width=0.89\textwidth{}]{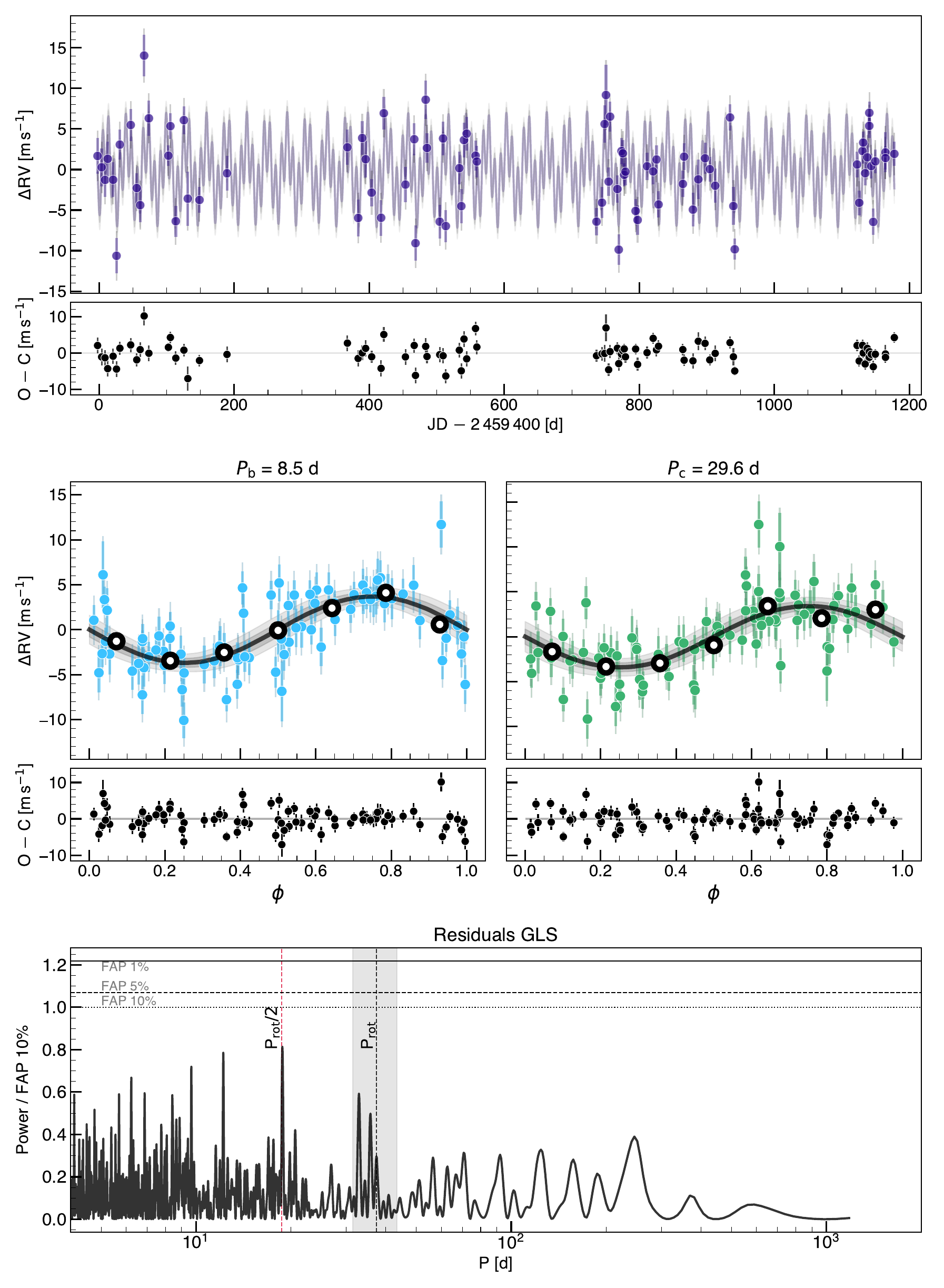}
\caption{\sbart\ RV time series (\textit{top} panel, purple dots) and phase-folded for both planets separately (blue dots in \textit{middle-left} panel for \planetb, and green dots in the \textit{middle-right} panel for \planetc). The quadratic trend, and the model of the other planet when plotted in phase (\textit{middle} panels), are subtracted from the RVs. The best model is shown as a solid line, with shaded background regions representing 1- and 2$\sigma$ confidence intervals. Error bars show the $\sigma_{\rm RV}$ (darker and thicker), and the $\sigma_{\rm jit}$ (lighter and thinner). The residuals of the model are shown in the panels below. Binned measurements in the $middle$ panels are shown as black circles. \textit{Bottom:} GLS periodogram from the residuals of the model. The grey shaded region indicates the stellar rotational period within 1$\sigma$, and its half is shown in dashed red vertical line. False alarm probabilities are indicated with horizontal black lines.}
\label{fig:fitrv}
\end{figure*}

The prior and marginalized posterior distributions for the selected model are shown in Table~\ref{tab:F7} (see also the corner plot in Fig.~\ref{fig:E6}). Also in Table~\ref{tab:F7}, we summarize the planetary properties inferred from the posteriors for \planetb\ and \planetc. In Fig.~\ref{fig:fitrv}, the top panel shows the median shape of the highest evidence model (2p1c2c) as a function of time, the middle panel shows the phase-folded curves for both orbital periods, and the bottom panel gives the residuals GLS periodogram. As we show in the latter, no additional significant periodicities were found. For this reason, models of higher complexity are not justified (3p1c2c3c or 3p). Indeed, we checked that the MCMC with three Keplerians did not converge to any solution but found over-densities in the posterior distributions close to 18.8\,d ($\sim P_{\rm rot}$/2) and 12.2\,d periodicities.
In Fig.~\ref{fig:E7} we display the RV residuals (after subtracting each planet, and both of them) against the activity indicators. The absence of correlation in all of the cases (Pearson and Spearman coefficients are below 0.25 in all cases) reinforces that none of the two detected signals are caused by activity. 

Although the RV time series do not show any signal at the stellar rotation period, the presence of the (yet non-significant) periodicity at around $P_{\rm rot}$/2 in the residuals encouraged us to test a model that includes a Gaussian processes (GP). We modeled the RV time series with two circular Keplerians (2p1c2c), a quadratic trend and the GP. We jointly modeled the CaIRT$_{\rm 2}$ time series with an independent quadratic trend and the GP to inform shared hyper-parameters (detailed below) with the RVs. This way, the CaIRT$_{\rm 2}$ serves as an activity proxy, since together with the CaIRT$_{\rm 1}$, it is one of the only indicators showing a clear signal at the rotation period as inferred from the \logRHK{} in Sect.\,\ref{sec:act} (see Fig~\ref{fig:actind}). We implemented the GP using a quasi-periodic kernel (e.g., \citealt{haywood14}; \citealt{rajpaul15}) through the \texttt{george}\footnote{Available at \url{https://george.readthedocs.io/en/latest/}} package (\citealt{2015ITPAM..38..252A}), which takes the form:

\begin{equation}\label{eq:qp}
    \Sigma_{ij} = \eta_1^2\,{\rm exp}\left[ -\frac{(t_i - t_j)^2}{2\eta_2^2} - \frac{2 {\rm sin}^2\left(\frac{\pi (t_i - t_j)}{\eta_3}\right)}{\eta_4^2} \right] + (\sigma_{i}^2 + \sigma_{\rm jit}^2)\, \delta_{ij}.
\end{equation}

Three of the hyper-parameters are shared between the RV signal and the proxy (aperiodic timescale, $\eta_2$; correlation period, $\eta_3$; and periodic scale, $\eta_4$), while the other two parameters corresponding to their GP amplitudes are independent ( $\eta_{1,~\rm{RV}}$ and $\eta_{1,~\rm{PX}}$). The proxy requires four additional parameters (a constant, $\gamma_{\rm{PX}}$; a linear and a quadratic term, $m_{\rm PX}$ and $q_{\rm PX}$; and an additional noise, $\sigma_{\rm jit,\,PX}$). In Eq.\,\ref{eq:qp}, $\sigma_{i}$ refers to the observed uncertainty and $\sigma_{\rm jit}$ to the jitter, both associated with the RVs or the proxy in each case. We chose uniform distribution for $\gamma_{\rm{PX}}$, $m_{PX}$, $q_{PX}$ (ranges equivalent to the ones set for $\gamma$, $m$, and $q$ respectively), $\eta_{1,~\rm{RV}}$, $\eta_{1, ~\rm{PX}}$ (equivalent to the ranges set for $K$), and $\eta_2$ (from 1.5 times $P_{\rm rot}$ to 3 times the time span). The prior for $\sigma_{\rm jit,\,PX}$ is a modified log-uniform equivalent to that for the RVs and we chose a log-uniform for $\eta_{4}$ to explore several orders of magnitude. The only informed prior is $\eta_{3}$, for which we chose a Gaussian distribution centered in $P_{\rm rot}$ with an standard deviation of 6\,d and truncated from 1 to 200\,d.

In Table~\ref{tab:F8} we show the prior and posterior distributions for the 2p1c2c\,+\,GP model. The GP finds a solution that explains very well the behavior of the CaIRT$_{\rm 2}$ (see Fig.~\ref{fig:E8}), but finds no counterpart for such activity in the RV signal, providing a posterior for its amplitude truncated at zero in a confidence interval of 95\%. We note that compatible distributions are found for $\eta_{\rm 1,\,RV}$ when testing GP models informed with other proxys (dLW or FWHM) and for models with zero and one planet. Including the GP, the presence of two planets is also preferred over zero and one planet based on the Bayes Factor. We thus concluded that the activity has no measurable effect on the RVs. We note that although the $P_{\rm rot}$ periodicity is not found in the RV time series, it is possible that the signal at 18.7\,d is actually its first harmonic. This behavior of the activity, with the first harmonic of the rotation period being more relevant than the rotation period itself has also been found in other systems (e.g., \citealt{dumusque17, suarez-mascareno17, georgieva23}). Since the signal is not significant, it is not expected to cause any effect in the planetary parameters here inferred.

\subsection{Search for transits in the TESS light curve}
\label{sec:TESSanalysis}

\begin{figure*}
\centering\includegraphics[width=1.0\textwidth{}]{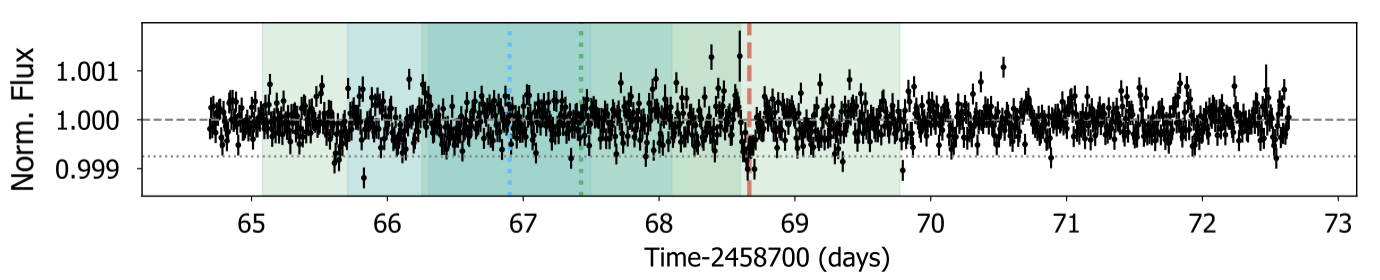}
\caption{TESS light curves of the first half of sector 17 (before the down-link gap) for \host{}. The shaded regions represent the 68.7\% (light) and 95\% (dark) confidence intervals for the conjunction time of \planetb{} (blue) and \planetc{} (green) according to the RV analysis in Sect.~\ref{sec:RVanalysis}. The vertical dotted lines correspond to the median of the expected transit time for each planet, while the vertical dashed red line indicates the location of the detected transit-like feature. The horizontal dashed line corresponds to 1 and the horizontal dotted line corresponds to the depth of a 2~\Rearth{} planet, for reference.}
\label{fig:TESSphot}
\end{figure*}

We used the PDCSAP photometry from the two sectors observed by the TESS mission (see Sect.~\ref{sec:tess-data}) on \host{}. This instrumentally corrected photometry is flat enough, so we did not need to apply any additional detrending model. Based on the RV analysis from Sect.~\ref{sec:RVanalysis}, we propagated the expected transit times and their corresponding uncertainties taking into account the uncertainties in both the time of conjunction and the period. The shaded regions in Fig.~\ref{fig:TESSphot} (first half of sector 17) and Fig.~\ref{fig:E9} (the remaining of sector 17 and sector 57) mark the expected location of the propagated transit times for both planets at 68.7\% and 95\% confidence intervals along both sectors. 

A blind application of the \texttt{tls} (Transits Least Square, \citealt{heller19}) to each sector independently suggests a transit with a signal detection efficiency (SDE) of 7.1 in sector 17 while no transit detection in sector 57. Interestingly, the detected signal in sector 17 corresponds to a period of 8.57~d, which is very close to the inferred period for \planetb{} from the RV analysis. However, the transit times found (corresponding to ${\rm BJD}-2\,459\,800$ = 68.66, 77.24, and 85.81) do not match the ephemeris of the RV model, being beyond 3$\sigma$. Indeed, the second transit time lays within the downlink gap of the spacecraft and the third one matches a small dimming incompatible in depth with the first one. This, together with the non-detection of transits along sector 57, suggests that the detected transit does not correspond to planet b.

This \texttt{tls} solution is driven by a dimming at time BJD\,=\,2\,458\,768.66466~d, which is within about 1.0$\sigma$ from the expected location of the transit for \planetc{} according to the RV ephemeris (see Fig. \ref{fig:TESSphot}). We then explored the possibility that this dimming is caused by the transit of the outer planet instead. This possibility would also explain the absence of transits in sector 57, since the expected time for the transit in this second sector assuming the transit time from sector 17 and the period from the RVs would be at BJD\,=\,2\,459\,865.40625252; unfortunately, this lies within the downlink gap of TESS. Consequently, there is only one potential transit for \planetc\ among the two available TESS sectors. We caution against claiming a planetary origin of this photometric signal due to the fact that there is only a single transit-like dimming and also because its depth is still at the level of other wiggles in the TESS light curve. In Appendix \ref{sec:tess_plc}, we analyze the dimming to test the transiting \planetc{} scenario. We conclude that although these results are promising, there is still no sufficient evidence to claim for a planetary origin of this TESS feature with the data on hand. Additional observations are needed to unveil its origin.

\section{Discussion}
\label{sec:discussion}

\subsection{Confirmation of the planetary nature}
\label{sec:verification}

At the moment of writing, discussions among the exoplanet community are ongoing with respect to defining an Exoplanet Confirmation Protocol (ECP, Lillo-Box et al., in prep.), aimed at clarifying the minimum requirements for a detected signal to become a confirmed planet. Under the current scheme, this protocol is based on the accomplishment of three generic principles: 1) signal and model significance (first generic principle); 2) demonstrated origin of the signal (second generic principle); and 3) confirming that the signal comes from an object in the planetary-mass domain (third generic principle) to ensure it accomplishes the IAU working definition of extrasolar planets (see \citealt{lecavalier22}). The KOBE experiment, as an RV-driven exoplanet search survey, is committed to abide to this protocol. In this section, we summarize the evidence to classify the two signals reported around \host{} as "confirmed planets''.

\begin{figure}
\centering
\includegraphics[width=0.5\textwidth{}]{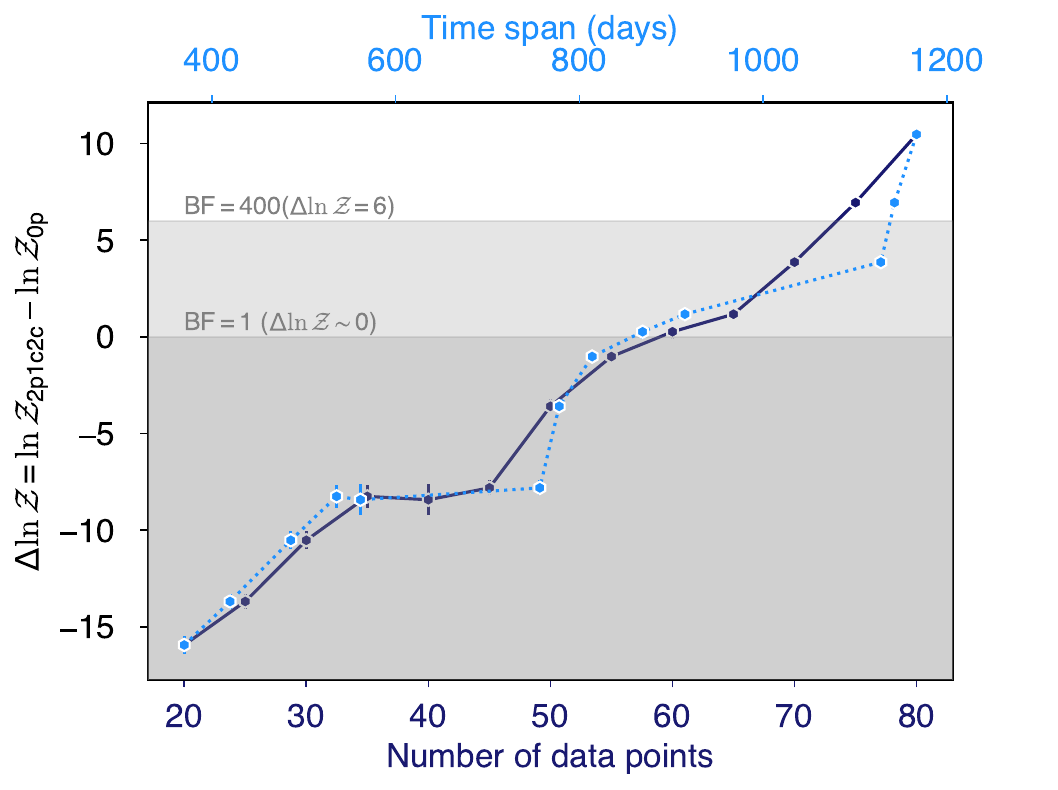}
\caption{Evolution of the difference between the evidence of the two-planet model and the zero-planet model as a function of the number of data points (dark blue solid line and symbols, corresponding to the \textit{lower} X-axis) and against the time from the first observation (light blue dotted line and symbols, corresponding to the \textit{upper} X-axis). The regimes of $\Delta \ln{\mathcal{Z}} < 6 $ (or BF = 400) and $\Delta \ln{\mathcal{Z}} < 0 $ (BF = 1) are shown as dark and gray shaded regions, respectively.}
\label{fig:EvidenceEvolution}
\end{figure}

\paragraph{Significance. }
 Throughout this paper, we  show that the RV signals of the two planets, \planetb{} and \planetc{}, reach a significant level, with the key parameter of the semi-amplitude of the RV converging to values above a 6$\sigma$ threshold ($K_{\rm b}=3.75\pm0.48$ ms$^{-1}$ and $K_{\rm c}=3.50\pm0.45$ ms$^{-1}$). In both cases, taking into account the stellar properties derived in Sect.~\ref{sec:stellarprop}, the amplitude of the signals are compatible with them being in the planetary mass domain, corresponding to minimum masses of $m_{\rm b}\sin{i_{\rm b}}$\,=\,\mb\,\Mearth{} and $m_{\rm c}\sin{i_{\rm c}}$\,=\,\,\mc~\Mearth. Regarding the significance of the model, the log-evidence of the two-planet model is favored over the one-planet hypothesis with respect to the commonly agreed significance threshold ($\Delta \ln\mathcal{Z}_{\rm 2p1c2c-1p1c}=$\,7.9). The evidence of these two-planet model against the null-hypothesis is very consistent in time, with a steady increase in the $\Delta \ln\mathcal{Z}_{\rm 2p1c2c-0p}$ metric over the course of the observations (see Fig.~\ref{fig:EvidenceEvolution}). Additionally, consistency is also seen in the evolution of the Bayesian GLS (BGLS, \citealt{mortier14, mortier17}) for both periodicities as new measurements are added (see the stacked BGLS in Fig.~\ref{fig:GLSevolution}). In Appendix~\ref{app:ev_K}, we also show the evolution of the RV semi-amplitudes for both periodicities, demonstrating that they are consistent throughout the four observed seasons. Hence, the signals from both planets satisfy the first generic principle of the ECP. 

\paragraph{Origin of the signal. }
The \gaia{}  renormalized unit weight error (RUWE, which evaluates the behavior of the center of light, see \citealt{lindegren18}) for \host{} is 1.057; hence, \gaia{} is able to provide a solid astrometric solution to this star, suggesting that no additional sources are present nearby (<\,1.4, \citealt{lindegren18}). Moreover, other metrics in {\em Gaia} DR3 help rule out long-term systemic RV variability, such as its uncertainty in the median of the epoch or similar criteria (see e.g., \citealt{katz23}) and the fraction of double transits (which evaluates the likelihood of an unseen stellar companion; e.g., \citealt{holl23}).  Also, the \gaia{} DR3 catalog does not include any additional sources within 10 arcsec. From the spectroscopy, the cross-correlation functions obtained by our \shaq\ pipeline do not show any additional sharp components that could highlight the presence of a companion blended within the CARMENES fiber aperture. Additionally, the bisector span and FWHM time series do not show correlation with the RVs (see Fig.~\ref{fig:actind}) that could be attributed to the RV variations being due to another star or a blended binary in a hierarchical triple with \host{}. All in all, the RV signal from both planets can be clearly attributed to \host{} and so we are in a position to fulfill the second generic principle of the ECP. 

\begin{figure}
\centering
\includegraphics[width=0.5\textwidth{}]{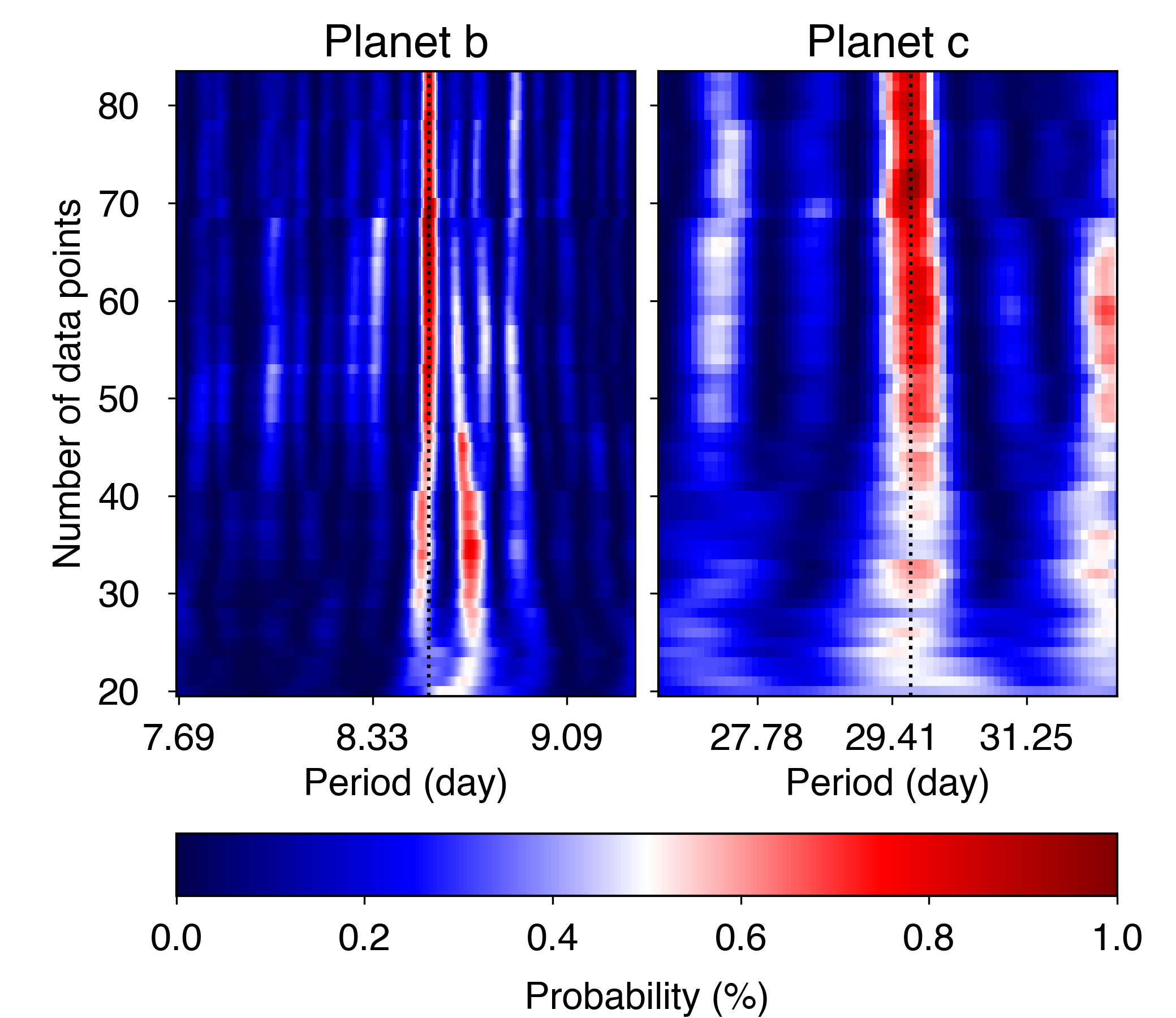}
\caption{Evolution of the BGLS periodogram as new measurements are added. \textit{Top} panels show this evolution for \planetb{} (\textit{left}) and \planetc{} (\textit{right}). \textit{Bottom} panel shows the color map indicating the probability of the Keplerian origin.}
\label{fig:GLSevolution}
\end{figure}

\paragraph{Planetary mass domain.}
The low minimum masses inferred for both signals imply that for these to be beyond the planetary-mass regime (13~\Mjup{}, as defined by the working definition of exoplanet from the IAU), the orbital inclinations must be below 0.12$^{\circ}$ and 0.17$^{\circ}$ for \planetb{} and \planetc{}, respectively. Assuming no preference for the orbital orientation with respect to our line of sight, the probability of these ranges are 0.13\% and 0.19\%; alternatively, there is a 99.87\% and 99.81\% of probability for both planets to be within the planetary-mass domain. These probabilities are larger than the 99.73\% threshold to consider a RV-only signal as a confirmed planet, thus both satisfying the third generic principle of the ECP.

\subsection{Warm super-Earths or sub-Neptunes}
\label{sec:warsemnep}

\begin{figure}
\centering\includegraphics[width=0.49\textwidth{}]{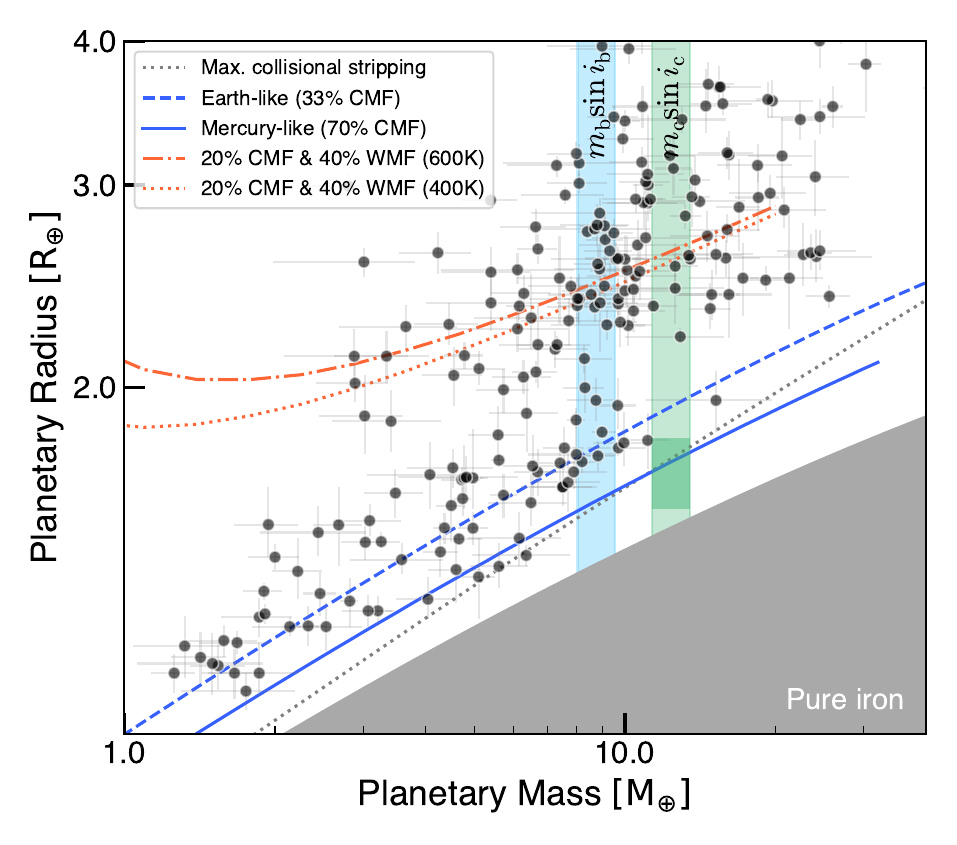}
\caption{Planetary mass-radius diagram. Vertical regions indicate the minimum mass within 1$\sigma$ of the planets detected around \host\ (blue for planet b, and green for c). Darker green region shows the 1$\sigma$ location for \planetc\ if we assume it causes the single-TESS transit (solely for illustrative purposes; see more in Sect.~\ref{sec:TESSanalysis} and Appendix~\ref{sec:tess_plc}). 
Different radius-mass models are shown as identified in the legend to guide the reader on the separated populations (blue lines trace tracks for super-Earths and orange lines for sub-Neptunes; \citealt{zeng19, aguichine21}). CMF and WMF refer to the core and water mass fractions, respectively.}
\label{fig:mr_pr}
\end{figure}

The minimum masses for both detected planets are compatible either with the super-Earth or sub-Neptune populations. As far as their orbital inclinations are above 35$^\circ$ (\planetb) and 55$^\circ$ (\planetc), their absolute masses would be compatible with the current super-Earth population (assuming as a limiting mass that of \object{WASP-84\,c} with 15.2$^{+4.5}_{-4.2}$\,M$_\oplus$). On the other side of the radius valley ($\gtrsim$1.8\,R$_\oplus$, \citealt{fulton17}), the sub-Neptune regime is less restrictive, ranging up to 25\,M$_{\oplus}$.

This degeneracy is shown in the radius-mass diagram of Fig.\,\ref{fig:mr_pr}, where the vertical colored lines display the 68.7\% confidence interval for the two planets minimum masses. The black dots show the exoplanet population, corresponding to confirmed planets with masses estimated from RVs, and relative errors for both mass and radius below 30\,\%\footnote{We discarded the planets HD~63433\,b and Kepler-107\,c since they are placed in the dense super-Earth domain but their mass estimations are not reliable according with their publications (\citealt{2024ApJS..272...32P}; \citealt{2023A&A...677A..33B}).} according with the NASA Exoplanet Archive table\footnote{\url{https://exoplanetarchive.ipac.caltech.edu/cgi-bin/TblView/nph-tblView?app=ExoTbls&config=PS}} (\citealt{2013PASP..125..989A}). For illustrative reasons, we included four possible interior models using the \texttt{mr-plotter} code\footnote{Available at \url{https://github.com/castro-gzlz/mr-plotter/tree/main}} (\citealt{castro-gonzalez23}) that track the sub-Neptune (orange) and super-Earth (blue) populations.

The darker green region in Fig.\,\ref{fig:mr_pr} indicates the \planetc{} location assuming that the TESS transit-like dimming corresponds with its transit (see Appendix~\ref{sec:tess_plc}, 1.69\,$\pm$\,0.12\,$R_{\oplus}$). Although this is very close to the maximum collisional stripping limit model (\citealt{2010ApJ...712L..73M}), such a scenario could be feasible. In such a case, \planetc\ would be compatible in terms of mass and radius with \object{WASP-84\,c} ($\rho$\,$\sim$\,2.0\,$\rho_\oplus$) and \object{TOI-1347\,b} ($\rho$\,$\sim$\,1.8\,$\rho_\oplus$, \citealt{2024ApJS..272...32P}), belonging to the densest super-Earths, with $\rho_{\rm c}$\,$\sim$\,2.3\,$\rho_\oplus$. However, while the ultra-short orbital period of the two known super-Earths (below 1.5\,d) can explain their small radius through photo-evaporation, this would not be the case for \planetc\ since it is low irradiated by the host star (S$_{\rm c}$\,$\sim$\,3.9\,S$_{\oplus}$, in a $\sim$30\,d orbital period). Future photometric observations of the target are therefore required to test this hypothesis.

\subsection{Sensitivity limits}
\label{sec:sensitivity}

To measure the sensitivity of the RV dataset to undetected Keplerian signals, we proceeded as in \cite{Standing2022}, \cite{Standing2023}, and \cite{John2023}. For this purpose, we first subtracted the solution from the two-planet model (the highest likelihood posterior sample with eccentricities $<$0.1) obtained when running \texttt{kima} with $N_p$ as a free parameters (see Sect.~\ref{sec:RVanalysis} and Table~\ref{tab:F6}). This sample has periods of $P_{\rm b}$\,$=$\,$8.539$\,d and $P_{\rm c}$\,$=$\,$29.625$\,d, semi-amplitudes of $K_{\rm b}$\,$=$\,$3.36$\,m\,s$^{-1}$ and $K_{\rm c}$\,$=$\,$3.34$\,m\,s$^{-1}$, and eccentricities of $e_{\rm b}$\,$=$\,$0.006$ and $e_{\rm c}$\,$=$\,$0.011$. To ensure that the residuals do not include any remaining of the signals, we run \texttt{kima} once again with $N_p$ as a free parameter. We confirmed that this yielded posterior samples favouring $0$ signals.

We then run \texttt{kima} on the residual data yet again following the process described in \cite{Standing2022} out to a maximum period of twice the time span of the data ($2\,000$~d). This time, the number of planetary signals was fixed to one ($N_p=1$). This yielded posterior samples with signals that are compatible with the data, though not formally detected.

\begin{figure}
\centering\includegraphics[width=0.5\textwidth{}]{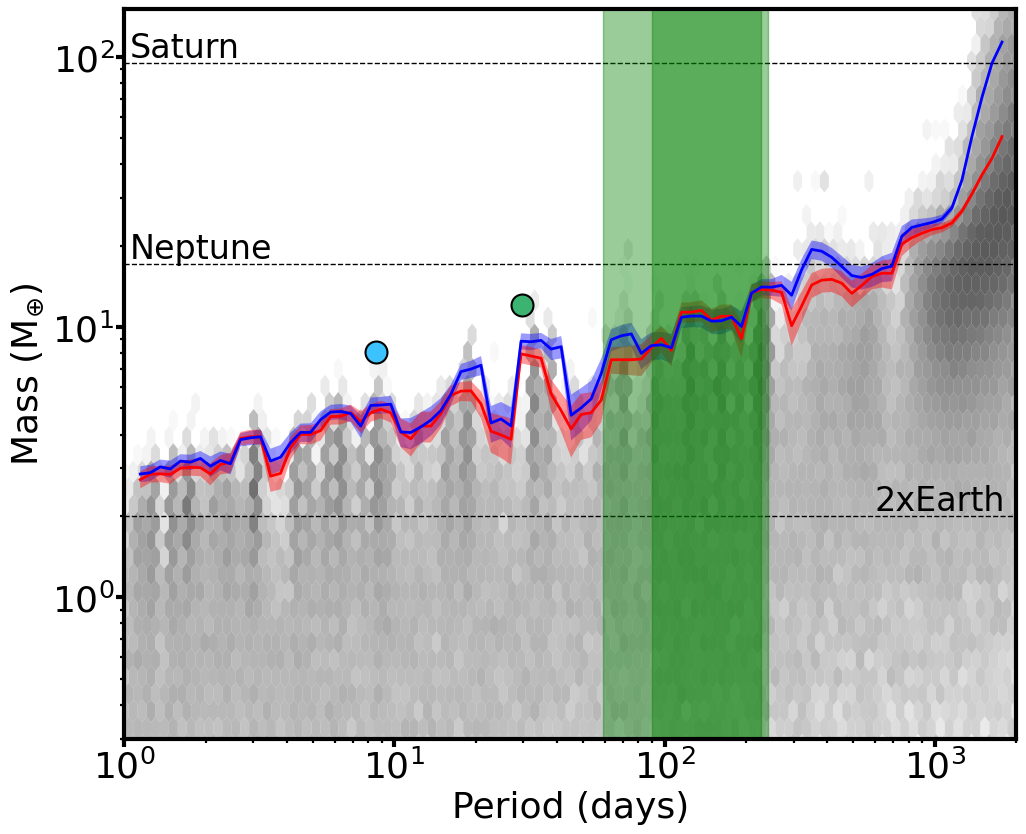}
\caption{Hexbin plot showing the posterior samples obtained from \texttt{kima} runs on the \host\ RV data with $N_p$ fixed to 1. The blue line shows the $99\%$ detection limit line, whereas the red line shows the same line computed on a subset of posterior samples with eccentricity $<0.1$. The uncertainties on these lines are illustrated by the faded lines of the associated color. The blue and green points represent \planetb\ and c, respectively. The extents of the optimistic and conservative HZ for the \host\ system are denoted by the light and dark green shaded regions.
}
\label{fig:senslim}
\end{figure}

In total, more than $400\,000$ posterior samples were obtained. These samples were then split into period bins, and the $99\%$ upper limit in each bin was calculated. This provided the detection limit which can be found in Fig.~\ref{fig:senslim}. Uncertainties on these limits were calculated as in \cite{Standing2023}. Signals above the blue detection limit line would have been detected in the initial \texttt{kima} run (checking there were no signals in the residuals). We can see that the red line using only samples with small eccentricities ($<0.1$) is lower on average than that of the blue line, which is why \cite{Standing2022} warns against the assumption of circular orbits when calculating detection limits.
Within the conservative HZ the \host\ dataset is sensitive down to masses of approximately $8.5~\rm{M_{\oplus}}$ at an orbital period of $\sim$\,90-105\,d. While, within the optimistic HZ we are sensitive down to masses of approximately $7.6~\rm{M_{\oplus}}$ at orbital periods of approximately $\sim$\,80\,d.

\subsection{Prospects for further characterization}
\label{sec:prospects}

\subsubsection{\gaia{} astrometry}
\label{sec:Gaia}

We computed the astrometric signatures for both \host\ planets by using the stellar and planetary properties from Tables~\ref{tab:stellar_param} and \ref{tab:F7}. We used the definition from \citealt{Perryman_2014}, with the signature being $\alpha=a_\star/d=(m_p/M_\star)\times(a_p/d)$, with the planetary semi-major axis $a_p$ in AU, and the system distance $d$ in pc. Hence, we obtained $\alpha_{\rm b}$\,=\,0.116\,$\pm$\,0.012 $\mu$as and $\alpha_{\rm c}$\,=\,0.403\,$\pm$\,0.041 $\mu$as.  
Even these estimated astrometric signatures are upper limits, assuming the most favorable co-planar face-on configuration ($\sin{i} \sim 1$) and assuming no overestimation in the mass, these signatures are too small even for the most precise astrometric measurements currently available, those published by \gaia{} in their Data Release 3 (\citealt{gaia22}). Since the expected \gaia{} end-of-mission parallax precision ($\sim 10\,\mu$as for 9\,$\leqslant G \leqslant$\,12, see \citealt{gaia18}) is still well above these estimated astrometric signatures, no evidence for the presence of \planetb\ and \planetc\ is expected from the astrometric technique.

\subsubsection{LIFE direct imaging}
\label{sec:LIFEsim}

Direct imaging is required to characterize non-transiting planets. However, the small angular separation ($\theta_{\rm b}$\,=\,2.9\,mas and $\theta_{\rm c}$\,=\,6.7\,mas) and masses of these two planets make them unsuitable for current instruments. For this reason, we used \texttt{LIFEsim}\footnote{Available at \url{https://lifesim.readthedocs.io/en/latest/index.html}} (\citealt{2022A&A...664A..22D}) to estimate the integration times that would be required to detect both planets with the LIFE mission (\citealt{2022A&A...664A..21Q}), a project based on nulling interferometry that is aimed at detecting warm worlds in the mid-infrared domain. 

We assumed the planets to emit as black bodies, for which we used their equilibrium temperatures collected in Table~\ref{tab:F7}. In terms of their size, we considered two possible scenarios for each planet (see Sect.~\ref{sec:warsemnep}): the super-Earths (SEs) and the mini-Neptunes (MNs). For \planetb\ we considered $R_{\rm b,SE}$\,=\,1.75\,$R_{\oplus}$ (Earth-like composition) and $R_{\rm b,MN}$\,=\,2.4\,$R_{\oplus}$ (20\% CMF, and 40\% WMF, for $T_{\rm eq,b}$ based on \citealt{aguichine21} models). Similarly, for \planetc, we used\ $R_{\rm c,SE}$\,=\,1.69\,$R_{\oplus}$ (Mercury-like composition based on the TESS dimming, see Appendix\,\ref{sec:tess_plc}) and $R_{\rm c,MN}$\,=\,2.5\,$R_{\oplus}$ (20\% CMF and 40\% WMF for $T_{\rm eq,c}$). We included all the astrophysical noise sources considered by \texttt{LIFEsim} (the stellar leakage, local zodiacal, and exozodiacal light) with a dust level of one zodi. Additionally, we tested the three instrumental scenarios available (optimistic, baseline, and pessimistic) varying the aperture diameter for each of the four-collecting telescopes (3.5\,m, 2\,m, 1\,m) and wavelength-range coverage (3-20\,$\mu$m, 4-18.5\,$\mu$m, and 6-17\,$\mu$m), with the spectral resolution fixed at 20. 

Following previous LIFE works (\citealt{2022A&A...664A..21Q}; \citealt{2022A&A...668A..52K}; \citealt{2023A&A...678A..96C}), we consider  a planet to be detected when the signal-to-noise ratio (S/N)  integrated over the wavelength range is above 7. Promisingly, all the simulations result in a detection. For \planetb\ the integration times span from 5\,min to 13\,h, and for \planetc\ from 10\,min to 44\,h, for the mini-Neptune \& optimistic and super-Earth \& pessimistic tests, respectively. We include in Table~\ref{tab:F9} the integration time for all the studied cases. These results place the \host\ planetary system as an interesting target for the LIFE mission.

\section{Conclusions}
\label{sec:conclusions}

In this work, we present the exoplanetary system \host, the first discovery of the KOBE experiment. It consists of two planets with orbital periods of \perb\,d (\planetb) and \perc\,d (\planetc). Over three consecutive years, we monitored the quiet star \host\ using the CARMENES spectrograph. These data enabled the detection of the planets with a 7$\sigma$ significance, which translates into a precise determination of their minimum masses with 11$\sigma$ significance ($m_{\rm b}\sin{i_{\rm b}}$\,=\,\mb\,M$_\oplus$ and $m_{\rm c}\sin{i_{\rm c}}$\,=\,\mc\,M$_\oplus$). Both planets, with minimum masses below that of Neptune, are consistent with either the super-Earth or mini-Neptune classifications.

We confirmed the planetary nature of \planetb\ and \planetc\ based solely on their RV data. This conclusion follows the principles of the ECP (Lillo-Box et al., in prep.), supported by the high significance of the signals, robust Bayesian evidence favoring the planetary scenario over simpler models, and the rejection of alternative non-planetary configurations with a probability above 99.7\%.

The outer planet orbits at a moderate distance from the star ($0.16071 \pm 0.00099$~au or a period of \perc\,d), albeit  with an insolation about 3.9 times that received by Earth from the Sun and; hence, it is located within the inner edge of the optimistic HZ. Nevertheless, our observational campaign offers a good coverage of this HZ, spanning three full seasons and the beginning of a fourth one, with each lasting around 200~d and with an average of 24 data points per season. A sensitivity analysis using the \texttt{kima} package ruled out the presence of planets with minimum masses above 8.5\,M$_\oplus$ within the HZ.

In Sect.~\ref{sec:TESSanalysis}, we explored a single dimming event detected by TESS in sector 17 that we cannot attribute to any of the two planets here discovered with the data in hand. Even if none of the \host\ planets transit, we conclude that this system could be fully characterized with future direct imaging space missions. Our analysis shows that the LIFE mission concept would require reasonable integration times to detect both planets: one hour for \planetb\ and four hours for \planetc,\ assuming a baseline configuration of the instrument and for the extreme scenario, where both planets are super-Earths rather than larger mini-Neptunes.

\section*{Data availability}

Figures E.2 - E.9 from Appendix \ref{sec:addfigs} and Tables F.3, F.4, F.6, F.8 - F.10 from Appendix \ref{sec:addtables} are also available in electronic form at \url{https://zenodo.org/records/14511891} and \url{https://zenodo.org//records/14516037}. The full version of Tables~\ref{tab:RVindicators} and \ref{tab:rv_sbart} are available in electronic form at the CDS via anonymous ftp to cdsarc.u-strasbg.fr (130.79.128.5) or via \url{http://cdsweb.u-strasbg.fr/cgi-bin/qcat?J/A+A/}.

\begin{acknowledgements}
We thank the anonymous referee for their revision of this manuscript that helped to improve its final quality.
We acknowledge the huge effort from the Calar Alto observatory personnel in running the KOBE experiment. In particular, Enrique de Guindos and Enrique de Juan for their outstanding work and help on the computing and data access side. We thank Rodrigo Fernando Díaz for the useful discussions on Bayesian statistics. We also thank Carlos Rodrigo Blanco (always in our memories) for his generous work in the development of the KOBE database.  
This Project has been funded by grants PID2019-107061GB-C61, PID2023-150468NB-I00 and MDM-2017-0737 by the Spanish Ministry of Science and Innovation/State Agency of Research MCIN/AEI/10.13039/501100011033.
J.L.-B. is partially funded by the NextGenerationEU/PRTR grant CNS2023-144309.
A.M.S acknowledges support from the Fundação para a Ciência e a Tecnologia (FCT) through the Fellowship 2020.05387.BD (DOI: 10.54499/2020.05387.BD). Funded/Co-funded by the European Union (ERC, FIERCE, 101052347). Views and opinions expressed are however those of the author(s) only and do not necessarily reflect those of the European Union or the European Research Council. Neither the European Union nor the granting authority can be held responsible for them. This work was supported by FCT - Fundação para a Ciência e a Tecnologia through national funds by these grants: UIDB/04434/2020, UIDP/04434/2020.
M.R.S. acknowledges support from the European Space Agency as an ESA Research Fellow.
E.M. acknowledges financial support through a "Margarita Salas" postdoctoral fellowship from Universidad Complutense de Madrid (CT18/22), funded by the Spanish Ministerio de Universidades with NextGeneration EU funds.
A.A. is supported by NASA’S Interdisciplinary Consortia for Astrobiology Research (NNH19ZDA001N-ICAR) under grant number 80NSSC21K0597.
J.C.M. acknowledges financial support from the Spanish Agencia Estatal de Investigaci\'on of the Ministerio de Ciencia e Innovaci\'on through projects PID2021-125627OB-C31, by “ERDF A way of making Europe”, by the programme Unidad de Excelencia María de Maeztu CEX2020-001058-M.
E.D.M. acknowledges the support from FCT through Stimulus FCT contract 2021.01294.CEECIND, and from the Ramón y Cajal grant RyC2022-035854-I funded by MICIU/AEI/10.13039/50110001103 and by ESF+.
A.B. was funded by TED2021-130216A-I00 (MCIN/AEI/10.13039/501100011033 and European Union NextGenerationEU/PRTR).
S.C.C.B. acknowledges the support from Fundação para a Ciência e Tecnologia (FCT) in the form of work of work through the Scientific Employment Incentive program (reference 2023.06687.CEECIND).
V.A. was supported by Funda\c{c}\~ao para a Ci\^encia e Tecnologia through national funds and by FEDER through COMPETE2020 - Programa Operacional Competitividade e Internacionalização by these grants: UIDB/04434/2020; UIDP/04434/2020; 2022.06962.PTDC.
S.G.S acknowledges the support from FCT through Investigador FCT contract nr. CEECIND/00826/2018 and  POPH/FSE (EC).
E.N. acknowledges the sup port by the DFG Research Unit FOR2544 “Blue Planets around Red Stars”.
The project leading to this publication has received funding from the Excellence Initiative of Aix-Marseille University - A*Midex, a French “Investissements d’Avenir programme” AMX-19-IET-013. This work was supported by the "Programme National de Planétologie" (PNP) of CNRS/INSU co-funded by CNES.
The KOBE archive and the VOSA code are both developed under the Spanish Virtual Observatory (svo.cab.inta-csic.es), project funded by MCIN/AEI/10.13039/501100011033/ through grant PID2020-112949GB-I00.
This research made use of \texttt{astropy}, (a community-developed core Python package for Astronomy, \citealt{astropy:2013,astropy:2018}), \texttt{SciPy} \citep{scipy}, \texttt{matplotlib} (a Python library for publication quality graphics \citealt{matplotlib}), \texttt{astroML} \citep{astroML}, \texttt{numpy} \citep{numpy}, and \texttt{dfitspy} \citep{dfitspy}.
This research has made use of NASA's Astrophysics Data System (ADS) Bibliographic Services, the SIMBAD database operated at CDS, the Exoplanet Follow-up Observation Program (ExoFOP; DOI: 10.26134/ExoFOP5) website, and the NASA Exoplanet Archive, the latter two are operated by the California Institute of Technology under contract with the National Aeronautics and Space Administration under the Exoplanet Exploration Program. This material is based upon work supported by NASA’S Interdisciplinary Consortia for Astrobiology Research (NNH19ZDA001N-ICAR) under award number 19-ICAR19\_2-0041.
\end{acknowledgements}

%
%

\bibliographystyle{aa} 
\bibliography{bibliography} 

\appendix
\onecolumn 
\section{Comparison of the RV extraction by different pipelines}
\label{sec:pipes}

\cite{silva22} presented the \sbart\ pipeline and compared the behavior of different RV extraction methods (CCF, classical template matching, and their semi-Bayesian template-matching approximation) with the stellar spectral-type (M-, K-, and G-dwarf stars). They found that \sbart\ provides a better precision (i.e., lower $\sigma_{\rm RV}$) than CCF-based pipelines for the three studied stellar spectral types. Besides, the \sbart\ precision as compared with those from classical template matching improves when going to earlier types than M-dwarfs (for K- and more notably for G-type stars, see Fig.~14 from their publication).

\begin{figure*}
\centering\includegraphics[width=0.9\textwidth{}]{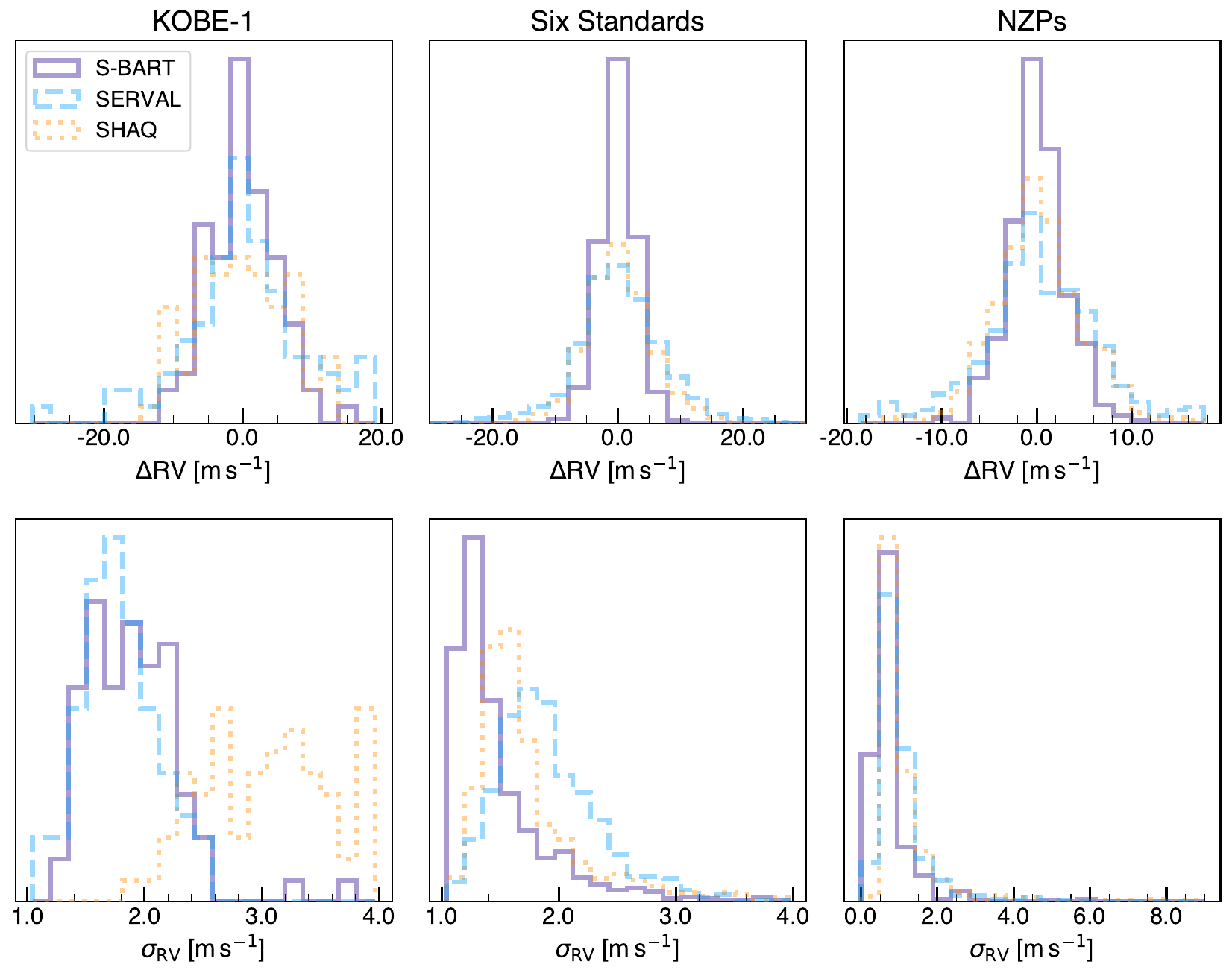}
\caption{RV distributions (\textit{top}) and their associated uncertainties (\textit{bottom}) for each pipeline (color code as shown in the legend of the top left panel). Columns correspond to \host\ (\textit{left}), the six KOBE standards (\textit{middle}), and the NZPs (\textit{left}).}
\label{fig:pipeshist}
\end{figure*}

In Fig.~\ref{fig:pipeshist} we compare the distributions of the RVs (top) and their uncertainties (bottom) extracted by the three pipelines used in this work (\sbart, \serval, and \shaq) for \host\ (left), the six KOBE standard stars (middle), and the NZPs (right). \host\ is a K7\,V star, a domain where classical template-matching algorithms (e.g., \serval) reach high precisions, being compatible with those from \sbart, and significantly better than CCF (e.g., \shaq) as can be seen in the bottom left panel of Fig.~\ref{fig:pipeshist}. Nonetheless, the spectral type of the standard stars used in KOBE range from  G6\,V to K0\,V. Consequently, \serval\ is not optimized to obtain the RVs of these six targets and, as a result, the RMS of the NZPs obtained with \serval\ double those of \sbart\ and have a 39\% lower precision (see middle and right columns from Fig.~\ref{fig:pipeshist}). As the \serval\ NZP correction is not optimal, if correcting with its NZPs there is a high dispersion of the final \host\ RVs (see top left panel from Fig.~\ref{fig:pipeshist}) that can hide the planetary signals (\perb\ and \perc\ d).

\begin{figure*}
\centering\includegraphics[width=\textwidth{}]{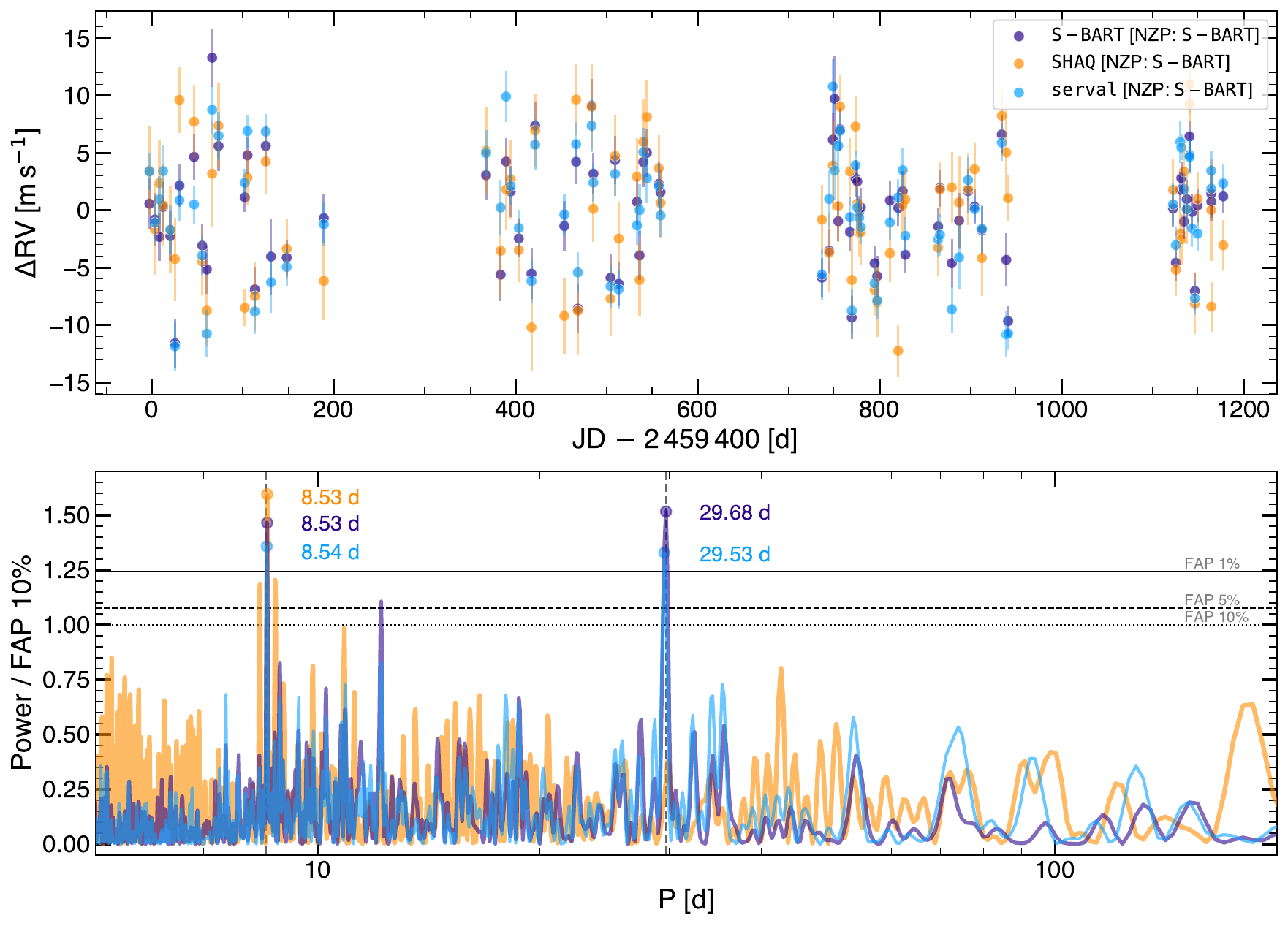}
\caption{Comparison of data from different pipelines. \textit{Top:} CARMENES RV time series extracted by \sbart, \serval\ and \shaq\ pipelines corrected with \sbart\ NZPs in all cases. \textit{Bottom:} Periodogram for the datasets shown above. The horizontal black lines indicate the different false alarm probabilities.}
\label{fig:sbartservalshaq}
\end{figure*}

As \sbart\ is the only pipeline extracting high quality RV measurements from the standards, we  corrected the CARMENES \host\ RVs from the three pipelines (\sbart, \serval, and \shaq) with \sbart\ NZPs. In Fig.~\ref{fig:sbartservalshaq} we show the GLS periodograms for these three time series, where the planet b periodicity (\perb\,d) is present for all of them, and the outer signal (\perc\,d) is only missed by \shaq\ probably as a result of its higher $\sigma_{\rm RV}$ (see again bottom left panel in Fig.~\ref{fig:pipeshist}). Therefore, we demonstrated that three different methodologies recover compatible time series. 

\vspace{0.5cm}
\section{The \perc-days periodicity}
\label{sec:moon}
 
\begin{figure*}
    \centering
    \includegraphics[width=\textwidth{}]{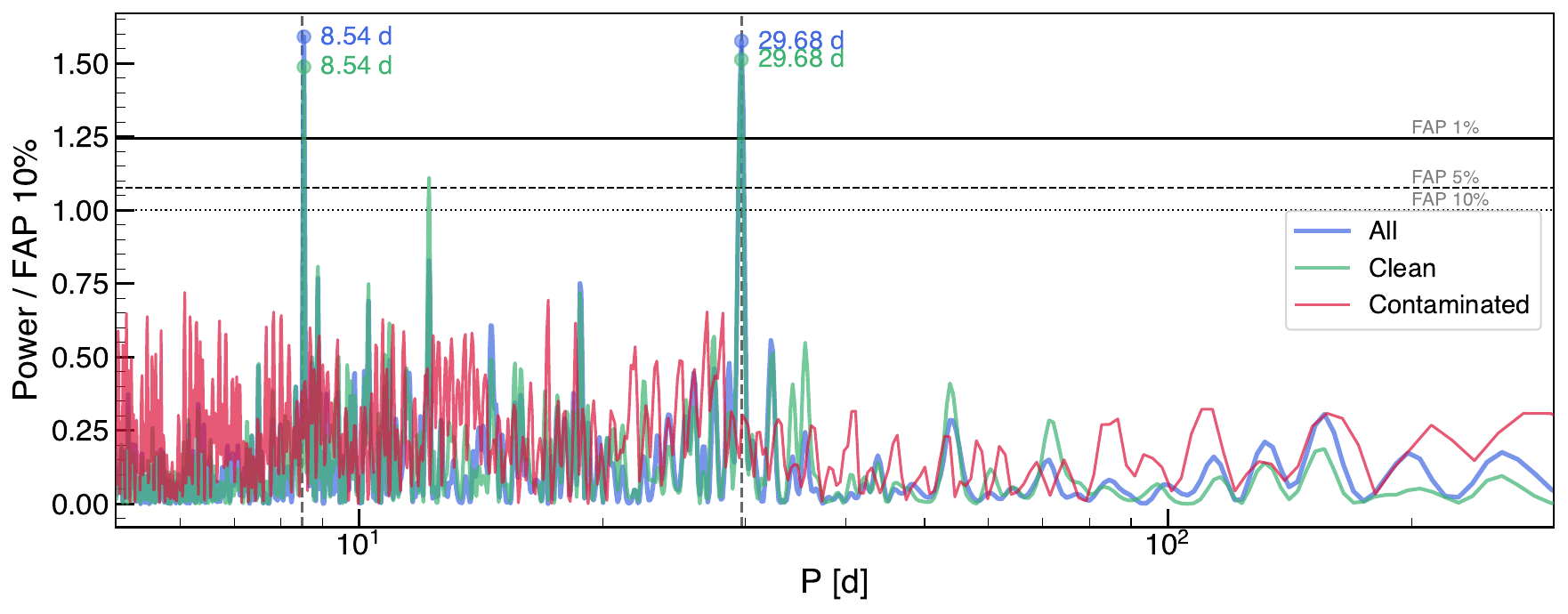}
    \caption{GLS periodogram of the full sample of RVs obtained for \host{} (blue line), the sub-sample of data points after excluding the possibly Moon-contaminated observations (green line) and the sub-sample of the affected data points (red line). The three false alarm probability level of 10\%, 5\% and 1\% are shown as black horizontal lines.}
    \label{fig:moon}
\end{figure*}

Given that the period of the external signal ($P_{\rm c} = 29.671_{-0.052}^{+0.050}$\,d) is close to the synodic period of the Moon (29.53~d), we checked the Moon separation and illumination of our observations. As detailed in Sect.~\ref{sec:obs}, we found that nine data points are potentially contaminated by the Moon illumination (including angular separations < $80^{\circ}$, Moon illumination > 40\% and BERV < 21\,km\,s$^{-1}$), hence we removed them from our dataset. In Fig.~\ref{fig:moon}, we show the periodogram of the complete dataset (blue line), the cleaned sample after removing the Moon-contaminated datapoints (green line) and the contaminated data points (red line). As shown, the periodicity at \perc~d remains in the cleaned sample and it is not present in the contaminated sample. The fact that the significance of the \perc~d signals decreases when removing this dataset is attributed to the fact that i) we are removing about 10\% of the data points, and ii) that the phases corresponding to the contaminated observations cluster in a region where there are few other uncontaminated observations. 

Another possible origin of this signal, also related to the synodic periodicity, is that the BERV correction is wrongly accounting for the Moon motion. If this would be the case, we would see similar periodicities in the RVs of other KOBE targets (whose BERV are calculated using the same procedure). This is not the case, and indeed, the only KOBE target showing a periodicity close to this synodic period is \host.

Additionally, the maximum impact that the
Moon contamination could have on \host\ RVs, based on its magnitude ($V \sim 10$\,mag) and spectral type (K7\,V), is below 0.3\,m\,s$^{-1}$ as estimated in \cite{cunha13} (see their Fig.~5), which is negligible for the CARMENES precision (> 1.5 m\,s$^{-1}$). Such an upper limit for the induced error is eleven times weaker than the signal found in this work ($K_{\rm c} = 3.50 \pm 0.45$\,m\,s$^{-1}$). All in all, our data and the analysis presented in this section supports that the origin of this \perc~d signal is different from the synodic period of the Moon. 

\section{TESS-CARMENES compatibility for a transiting \planetc}
\label{sec:tess_plc}

We analyzed the dimming found in Sect.~\ref{sec:TESSanalysis} assuming its origin from \planetc{} to extract the compatible planetary parameters. We used the \texttt{batman}\footnote{Available at \url{https://github.com/lkreidberg/batman?tab=readme-ov-file}} code (\citealt{batman}) to produce the planet transit model. Since we aimed at checking the compatibility of the transit signal with the RV solution for \planetc, we used a tight Gaussian prior on its orbital period based on the marginalized posterior distribution from Sect.~\ref{sec:RVanalysis}, and a uniform prior on the time of mid-transit including the whole first part of sector 17. We then imposed a uniform prior on the inclination allowing for large values not inducing transits, and a uniform prior on the planet radius of $\mathcal{U}(0,10)$~\Rearth. We estimated the quadratic limb-darkening coefficients using the \texttt{limb-darkening}\footnote{Available at \url{https://github.com/nespinoza/limb-darkening}} code from \cite{espinoza15} using the stellar parameters derived in Sect.~\ref{sec:stellarprop}, and used a Gaussian prior around these values with a 10\% width. We also used the stellar radius as a parameter with a Gaussian prior based on the results from Sect.~\ref{sec:stellarprop}. Finally, we added an offset parameter for the light curve and a $\sigma_{\rm jit}$ term with sufficiently broad priors. All these priors are specified in Table~\ref{tab:F10}. We sampled the parameter space using \texttt{emcee}, with four times as many walkers as parameters and 50\,000 steps per walker with the subsequent second phase as done in the RV analysis (see Sect.~\ref{sec:RVanalysis}). 
This analysis resulted in the convergence of the chains towards the transit detected in this sector despite using a conservative uniform prior on the time of conjunction. 

\begin{figure*}
\centering\includegraphics[width=0.49\textwidth{}]{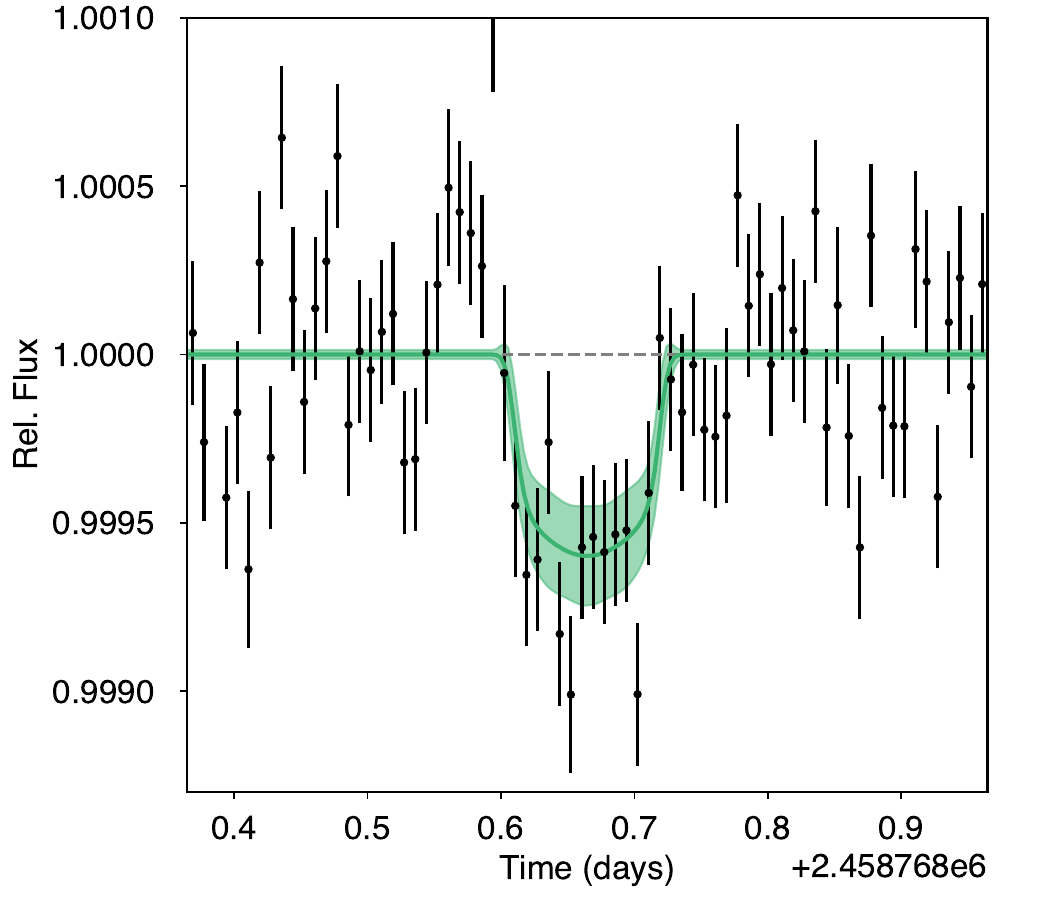}
\caption{Transit modeling of the feature detected in the TESS data from sector 17 that may be attributed to \planetc{}. The black symbols show the TESS 2-min cadence data from this sector. The green solid line corresponds to the median model from Appendix~\ref{sec:tess_plc} and the shaded region to models within the 68.7\% confidence interval.}
\label{fig:TESSphase}
\end{figure*}

The solution converged to a planetary transit with a radius of $1.69\pm0.12$~\Rearth\ and an orbital inclination of $89.219^{+0.047}_{-0.052}$ degrees (corresponding to an impact parameter of $0.760^{+0.036}_{-0.043}$), see Fig.~\ref{fig:TESSphase}. The transit duration of $2.79^{+0.17}_{-0.15}$~hours is within the expectations for a period like the one from \planetc{} without requiring a grazing configuration. The median and 68.7\% confidence intervals extracted from the marginalized posterior distributions of the inferred parameters, together with other derived parameters are shown in Table~\ref{tab:F10}.

By performing a joint analysis including the RVs and the TESS light curve, we further tested the compatibility of the signals. This time, we set the priors on the time of conjunction and period of the two-planet model as uniform in a wide range in order to neither bias the final result nor force the transit to be attributed to any of the RV-detected planets. We set the same period priors as for the RV-standalone analysis in Sect.~\ref{sec:RVanalysis} and the priors on the planet and stellar properties as in Sect.~\ref{sec:TESSanalysis}. We run a long 100\,000-step chain with a large number of walkers (10 times the number of free parameters) and sample the posterior distribution of the parameters using \texttt{emcee}. 

By the end of the run, none of the chains ended up in a configuration that would produce a transiting signal. However, the periods are well defined and converged to those from planets b and c. Only 0.3\% of the samples provided an impact parameter compatible with a transiting signal ($b< 1+R_p/R_{\star}$). All of them correspond to periods from \planetc{}. Those samples correspond to times of conjunction that would make the TESS observations miss the transit. Hence, the joint MCMC analysis definitively suggests that the current data do not prefer the transiting hypothesis for \planetc. 

We suggest several possible explanations for this result. One option is that the transiting signal corresponds to instrumental systematics or stellar noise, as despite being the deepest signal in the data, it is still compatible with the amplitude of other features in other parts of the light curve. Given that only one of such signals is observable in the TESS dataset, we have no further information to test this scenario. Second, even if the feature actually corresponds to a transit of \planetc{}, there is some discrepancy at the 1.0$\sigma$ level between the expected time of conjunction from the RV-standalone analysis and the location of the feature. This tension, together with the shallow and lonely feature, implies that the MCMC prefers to better model the RV data points than finding a compromise between the RVs and the potential transit signal, thus converging to non-transiting configurations. It might be possible that there are relevant transit timing variations accounting for that time discrepancy in case that \planetc{} forms part of a resonant chain.
Given this analysis, we remain cautious and do not claim that the detected signal is due to a transiting planet in the system. We note that TESS revisits this target in October 2024, during a period when \planetc\ is expected to transit.

\section{Evolution of the Keplerian signals}
\label{app:ev_K}

\begin{figure*}
    \centering
    \includegraphics[width=0.5\textwidth{}]{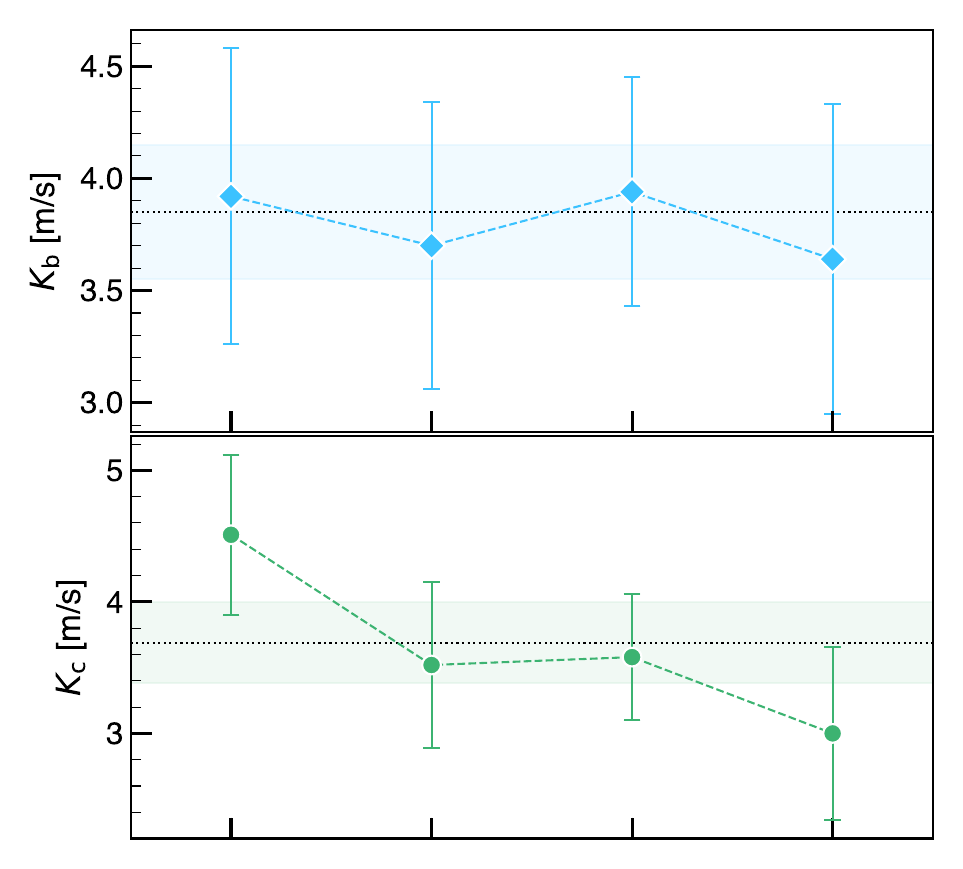}
    \caption{RV semi-amplitude evolution with the epoch for the signal induced by \planetb\ (\textit{top}) and \planetc\ (\textit{bottom}). Dotted horizontal line represent the mean of the RV semi-amplitude obtained when fitting all epochs with its 1$\sigma$ shown as a colored shaded region.}
    \label{fig:ev_K}
\end{figure*}

We studied the evolution of the RV semi-amplitudes of the two detected Keplerian signals as an additional test for their consistency. We first modeled the complete RV time series using the quadratic trend and the two Keplerians through the least-square method implemented in the DACE\footnote{\url{https://dace.unige.ch/dashboard/}.} web (\citealt{delisle16}). We note that the solution found from this method is compatible with the Bayesian approaches from Sect.~\ref{sec:RVanalysis}, but for this test we opted for a computationally cheaper analysis. We then split our dataset into four different chucks corresponding to the four visibility time spans from 2021 to 2024. These chunks contain individually an average of 20 data points per chunk. We then model each chunk separately with only the two RV semi-amplitudes as free parameters and fixing all the remaining orbital parameters from the previous solution. We show the evolution of the semi-amplitudes in Fig.~\ref{fig:ev_K}. 

Signals induced by stellar activity are known to be unstable with time, hence varying both their phase and amplitude (e.g., \citealt{mortier17}). This is not the case for any of the two signals found. In Fig.~\ref{fig:GLSevolution} we demonstrated that the signals are significant and their periodicities are stable with the number of measurements. In this appendix, we found that both signals are constant through the time span with results compatible for all epochs within their uncertainties. Therefore, the RV signatures cannot be related with activity and are more likely induced by planetary companions.

\section{Additional figures}
\label{sec:addfigs}

\begin{figure*}
\centering\includegraphics[width=0.94\textwidth{}]{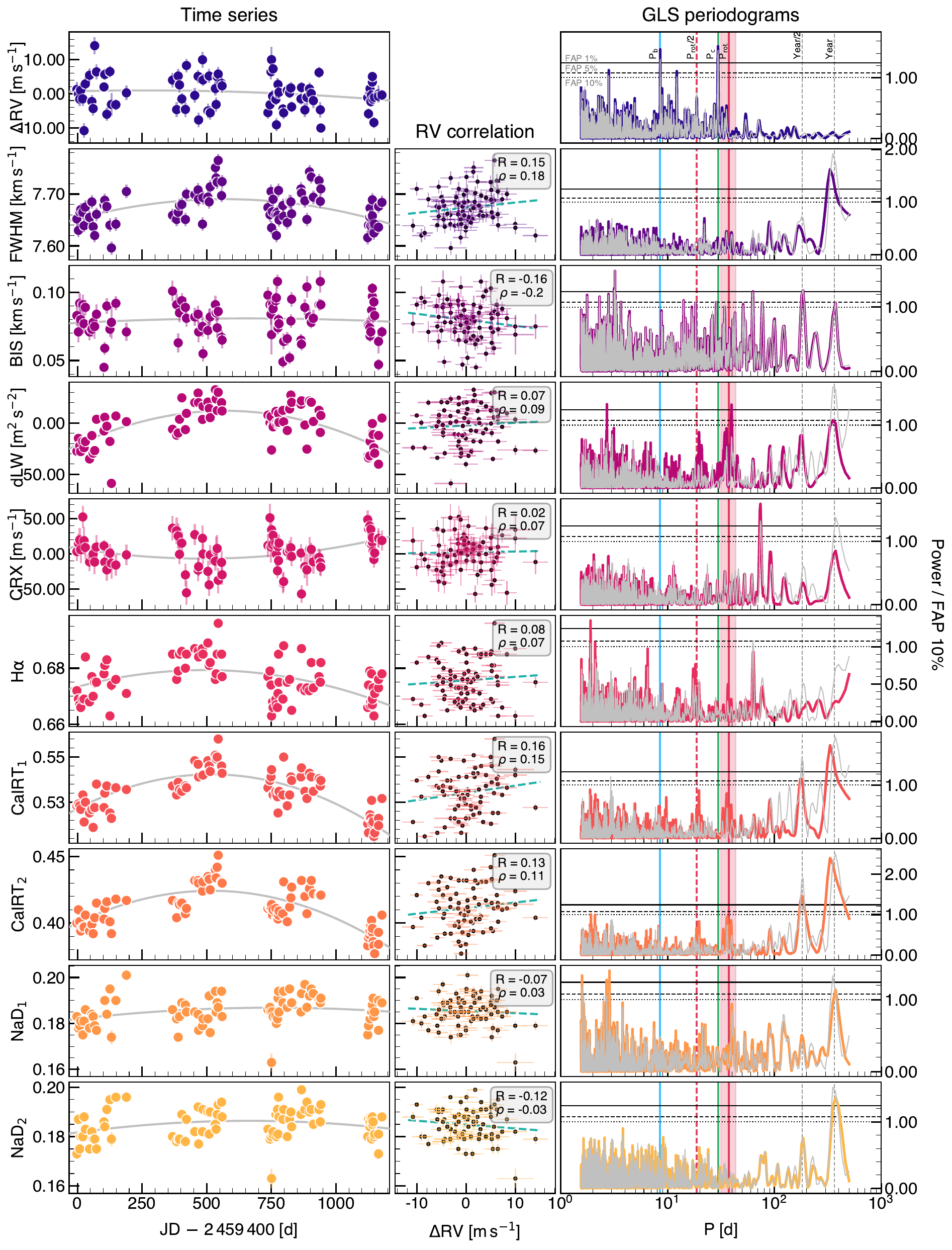}
\caption{CARMENES data. \textit{Left:} Time series for the \sbart\ RVs and the spectroscopic activity indicators. The grey line shows a quadratic trend fit. \textit{Center:} Activity indicator and RV correlation, with a linear fit shown in dashed blue line, and the correlation coefficients (R from Pearson, and $\rho$ from Spearman) indicated in the grey box. \textit{Right}: GLS periodogram to the time series of the same row. The grey line corresponds to the non-detrended time series, while colored lines correspond to the same dataset after subtracting the fitted quadratic function. Vertical lines indicate the periodicities of the planets (blue and green), the rotational period (red region, see Sect. \ref{sec:act}) and its half (dashed red line), and the duration of the year and its half (dashed grey lines). Horizontal lines show different false alarm probabilities as indicated in the top panel.}
\label{fig:actind}
\end{figure*}

\setcounter{figure}{1}
\begin{figure*}
\centering\includegraphics[width=0.45\textwidth{}]{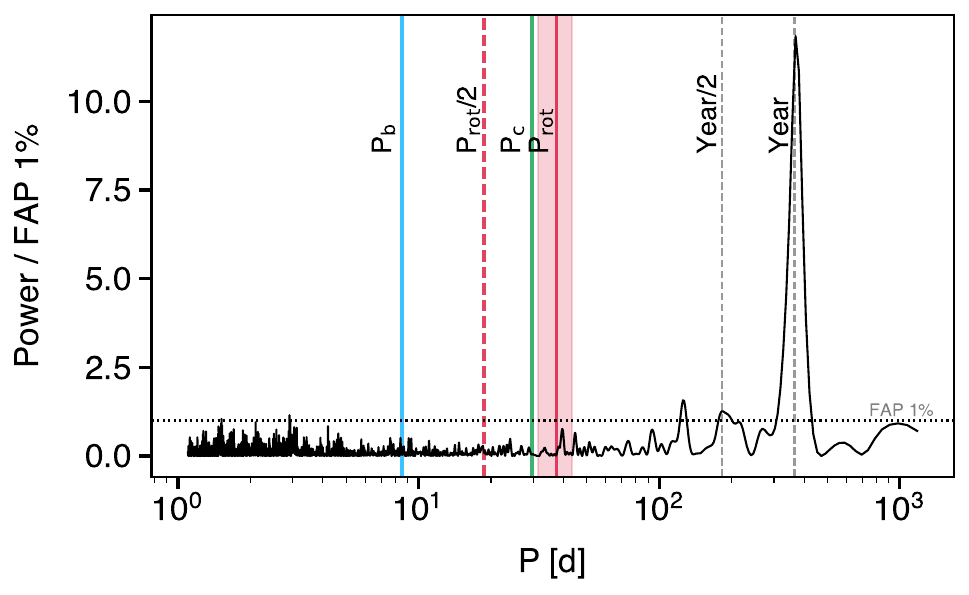}
\caption{Spectral window function of the CARMENES RV time series. See Sect. 2.1.}
\label{fig:E2}
\end{figure*}

\begin{figure*}
\centering
\includegraphics[width=0.45\textwidth{}]{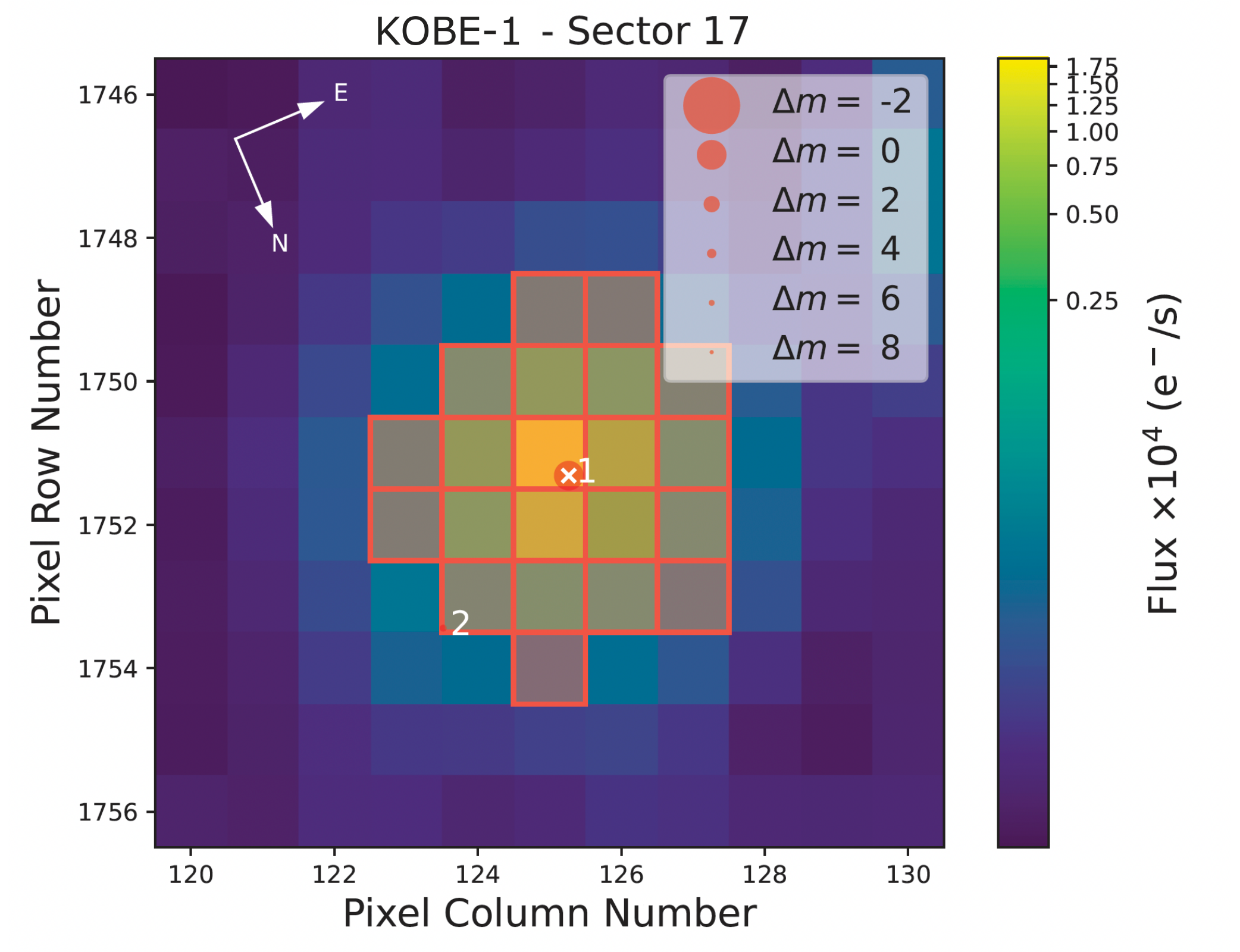}
\includegraphics[width=0.45\textwidth{}]{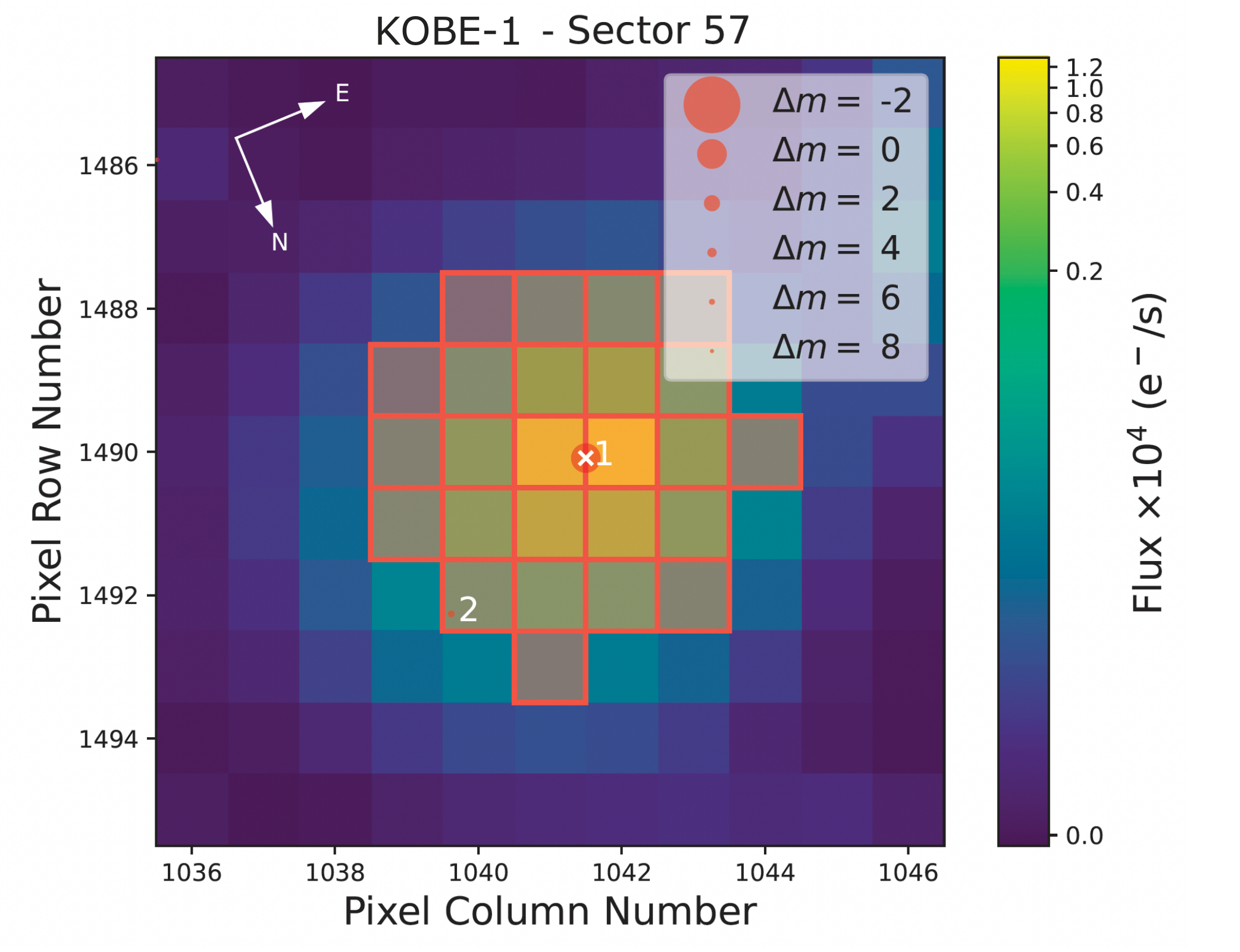}
\caption{TESS target pixel file (TFP) of \host\ for sectors 17 (\textit{upper} panel) and 57 (\textit{bottom} panel) using the \texttt{tpfplotter} algorithm. The white cross represents the position of the target, while the red dots are objects in the field detected by \gaia\ with delta $G$ < 8 mag. SPOC aperture is shown as red squares. See Sect.~2.3.}
\label{fig:E3}
\end{figure*}

\begin{figure*}
    \centering
    \includegraphics[width=0.45\textwidth]{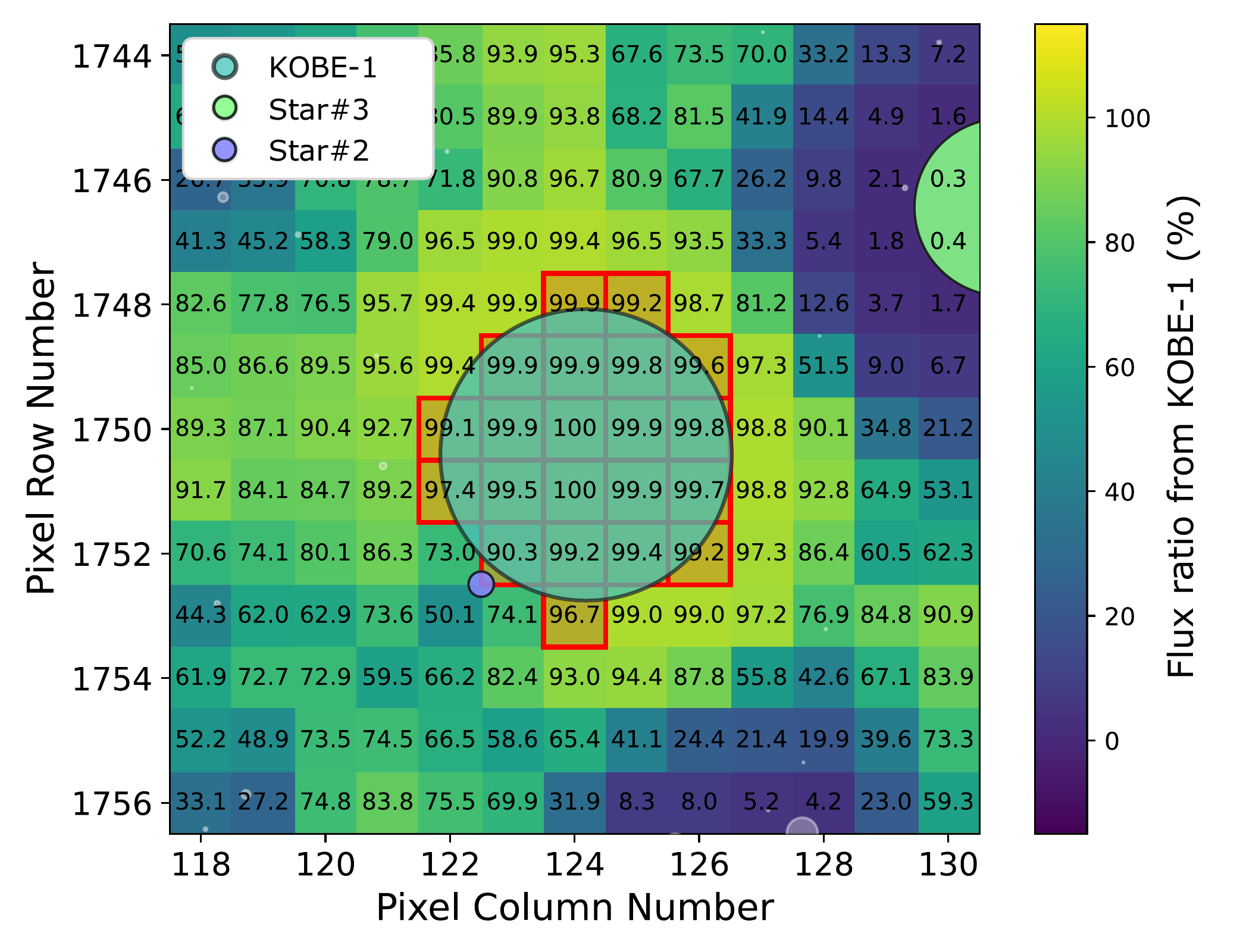}
    \includegraphics[width=0.45\textwidth]{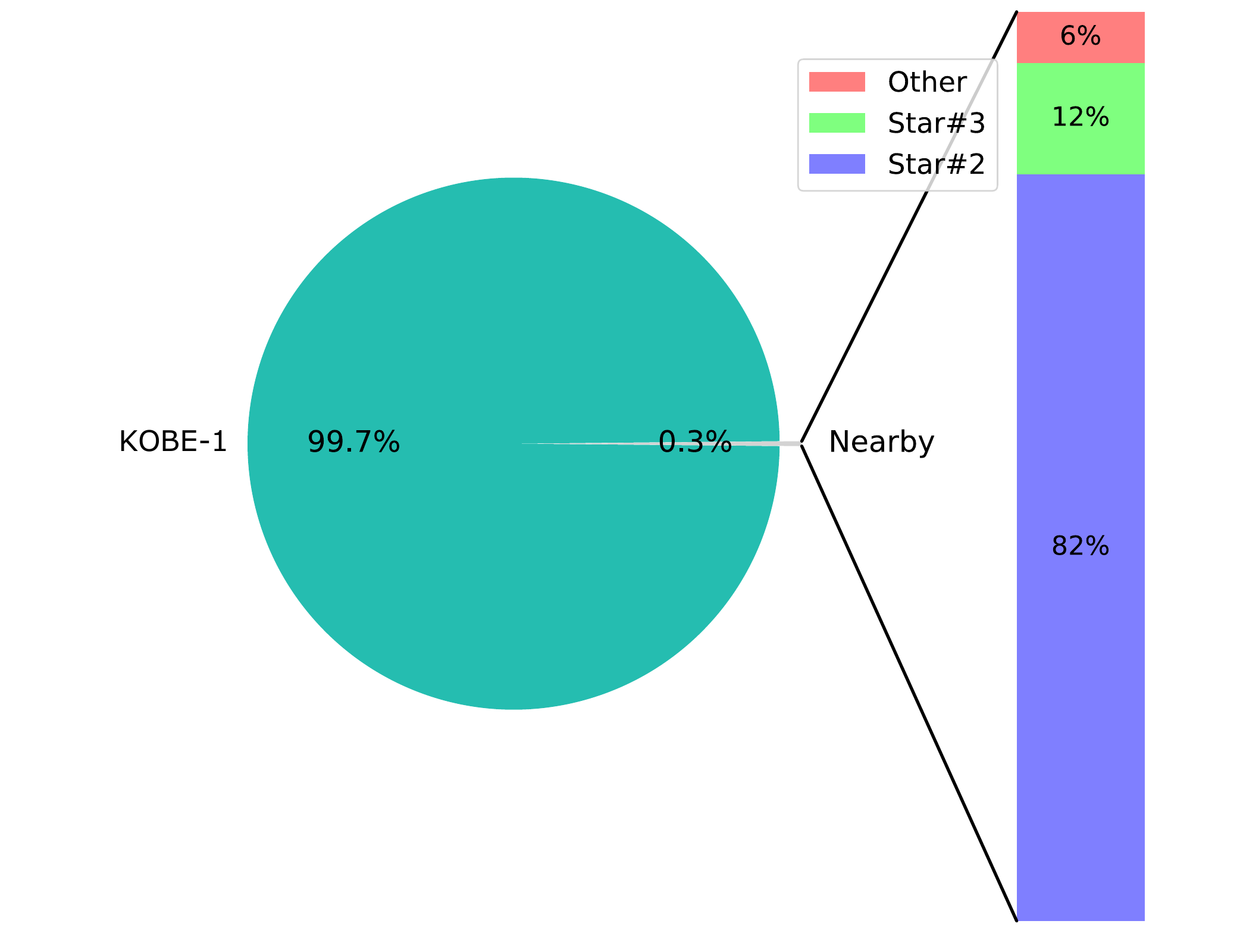}
    \caption{Nearby sources contributing to the TESS photometry of \host. \textit{Top}: Heatmap with the pixel-by-pixel flux fraction from \host\ in sector 17. The red grid is the SPOC aperture. The pixel scale is 21 arcsec $\rm pixel^{-1}$. The small-white dots represent all the \gaia{} sources, and the two sources that most contribute to the aperture flux are highlighted in purple and green as shown in the legend. The areas of the dots scale with the emitted fluxes. \textit{Bottom}: Pie chart showing the flux contributions to the SPOC aperture. This plot has been generated with \texttt{tess-cont}. See Sect.~2.3.}
    \label{fig:E4}
\end{figure*}

\begin{figure*}
    \centering
    \includegraphics[width=\textwidth]{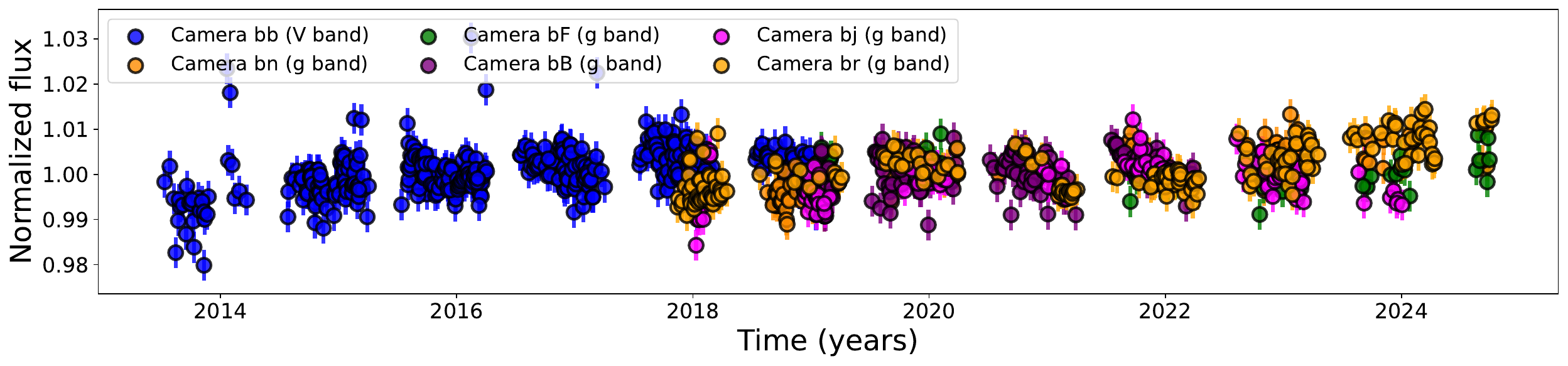}
    \includegraphics[width=\textwidth]{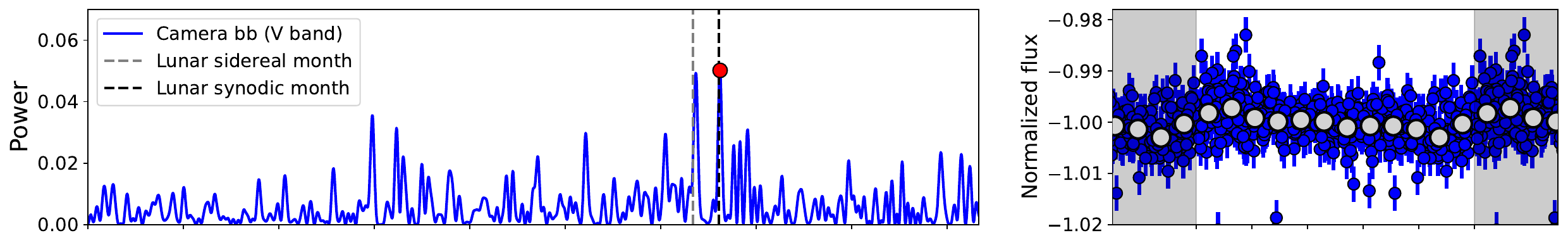}
    \includegraphics[width=\textwidth]{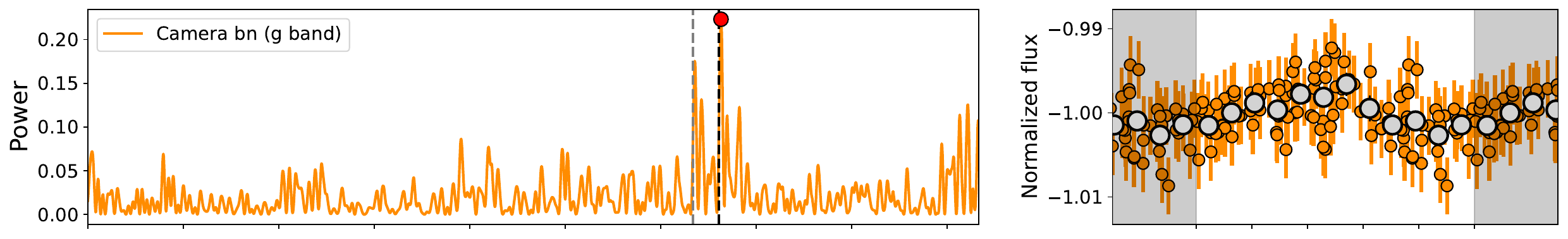}
    \includegraphics[width=\textwidth]{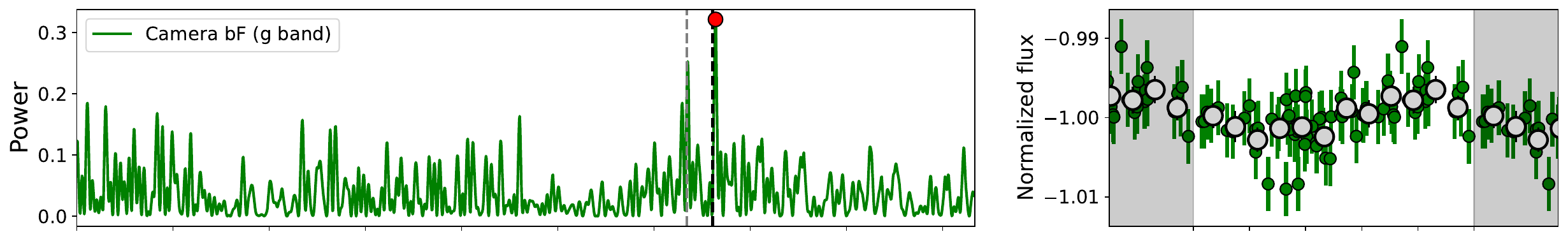}
    \includegraphics[width=\textwidth]{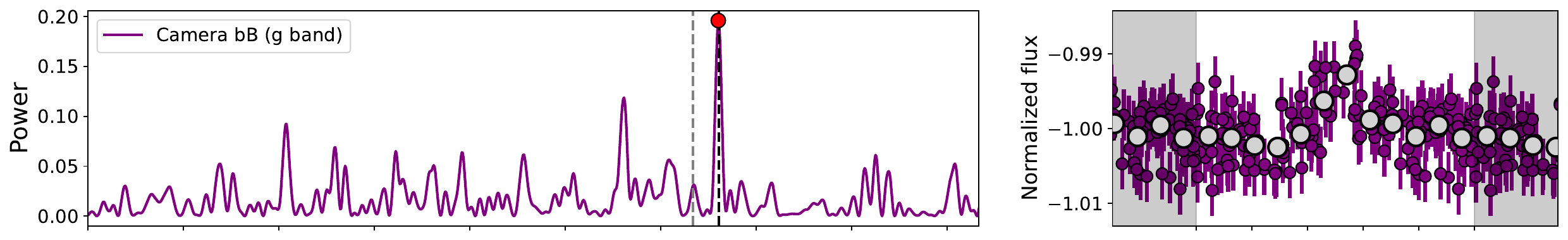}
    \includegraphics[width=\textwidth]{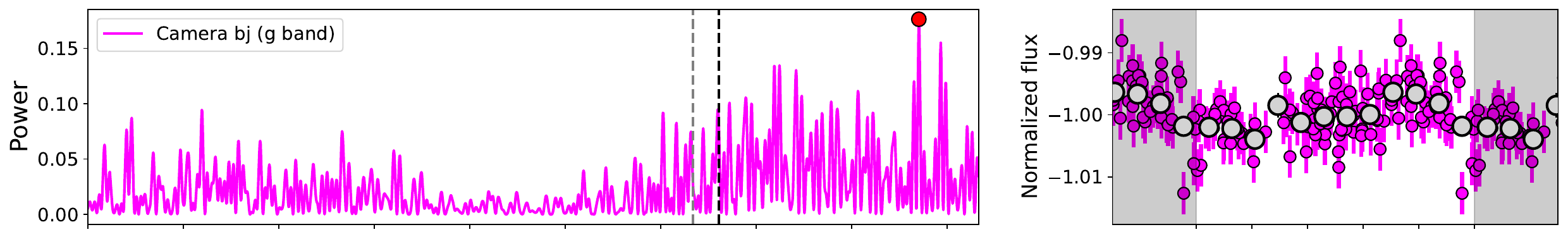}
    \includegraphics[width=\textwidth]{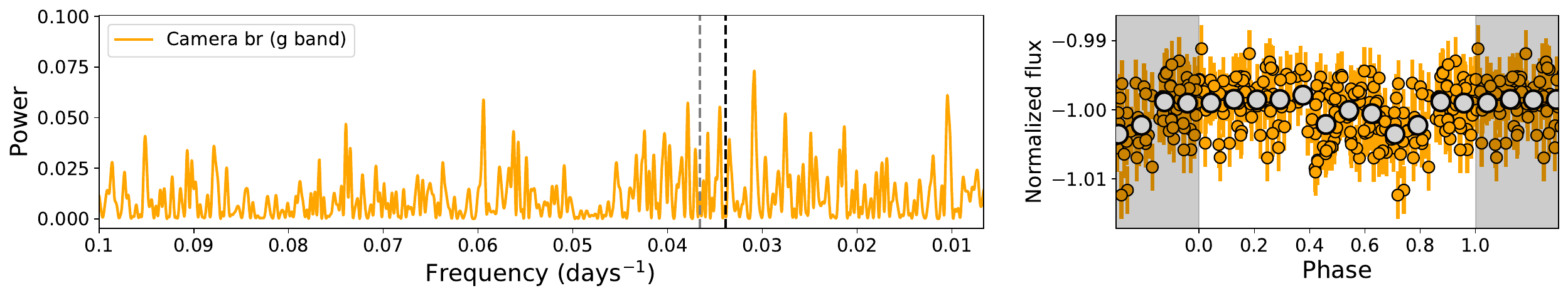}
    \caption{\textit{Top:} ASAS-SN photometric time series of \host{}. \textit{Left}: \texttt{GLS} periodograms of the ASAS-SN photometry corrected for a long-term linear trend. The vertical dashed lines indicate the lunar synodic (black) and sidereal (grey) month.  The horizontal dotted lines indicate the 10$\%$ (orange), 1$\%$ (blue), and 0.1$\%$ (green) FAP levels. \textit{Right:} Phase-folded light curves to the maximum power period obtained by the \texttt{GLS} periodogram (indicated with a red dot). The gray data points correspond to a 2.5-day binning. See Sect. 2.3.}
    \label{fig:E5}
\end{figure*}

\begin{figure*}
\centering\includegraphics[width=1.0\textwidth{}]{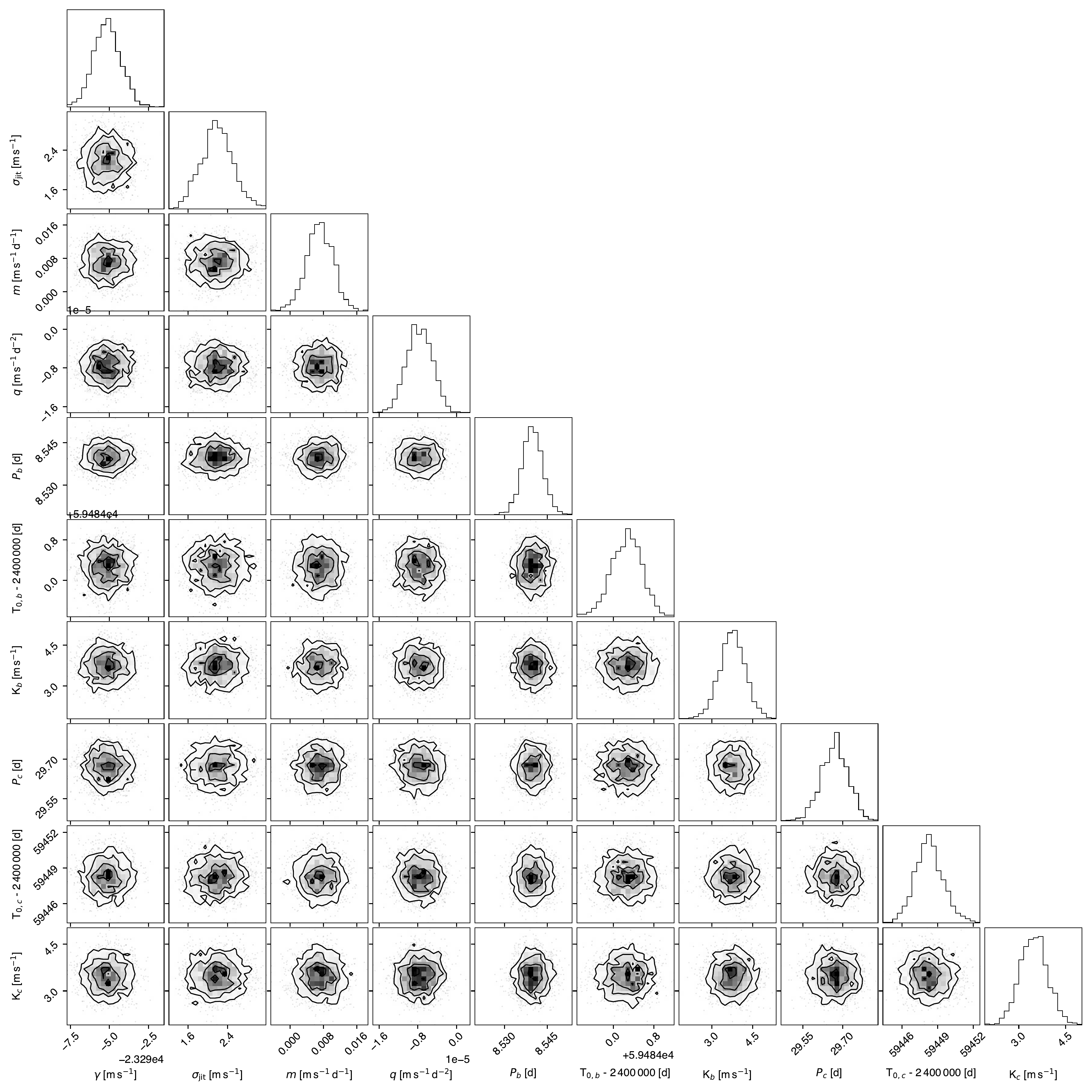}
\caption{Corner plot for the preferred RV model (2p1c2c, two planets in circular orbits). See Sect. 4.1.2.}
\label{fig:E6}
\end{figure*}

\begin{figure*}
\centering\includegraphics[width=0.94\textwidth{}]{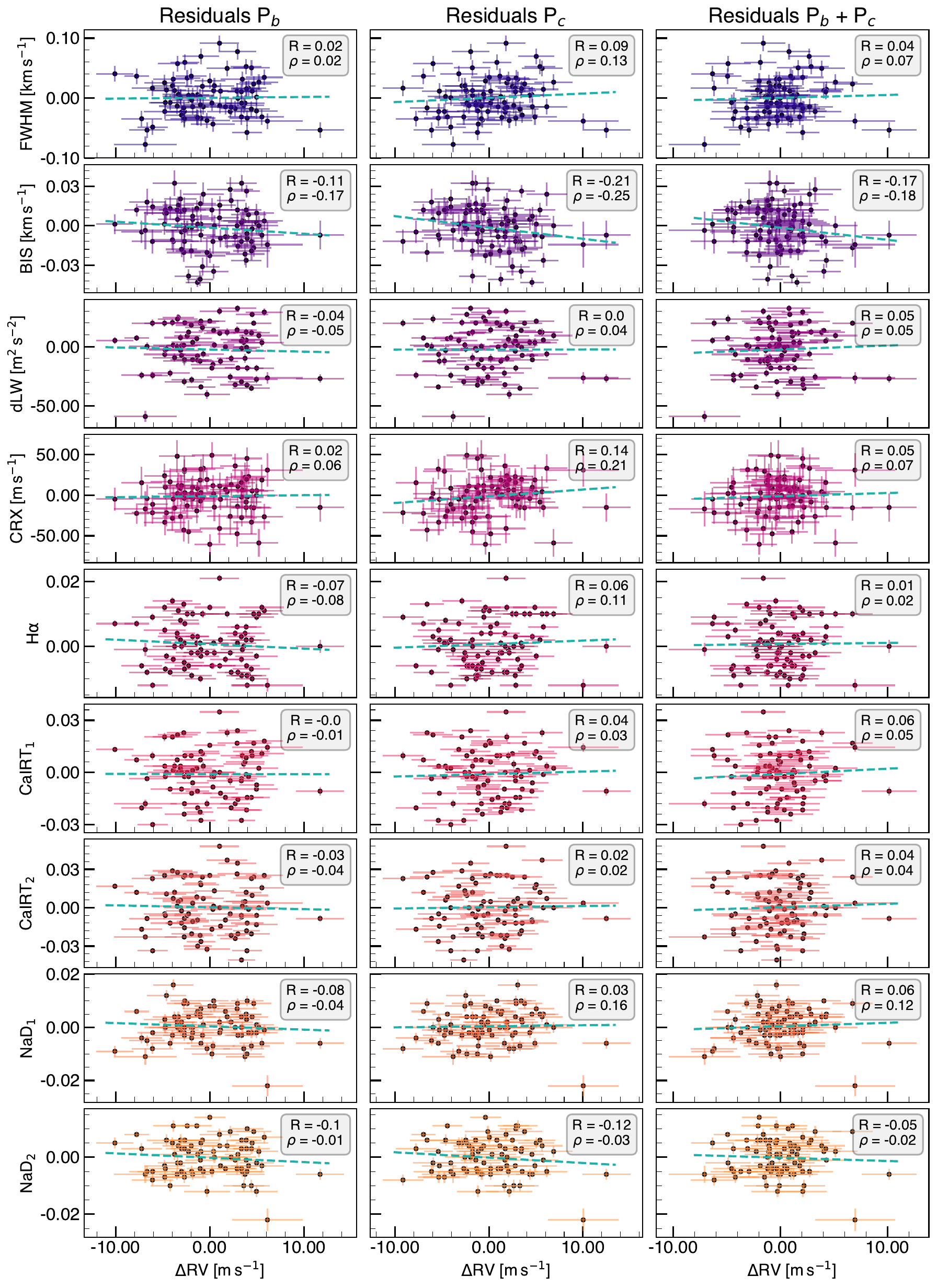}
\caption{Correlation for the RV residuals (after subtracting planet b in the \textit{left} column, planet c in the \textit{middle}, and both planets in the \textit{right}) with the activity indicators. Blue line is the linear regression, and in the grey box are shown the correlation coefficients (R from Pearson, and $\rho$ from Spearman). See Sect. 4.1.2.}
\label{fig:E7}
\end{figure*}

\begin{figure*}
\centering\includegraphics[width=1.0\textwidth{}]{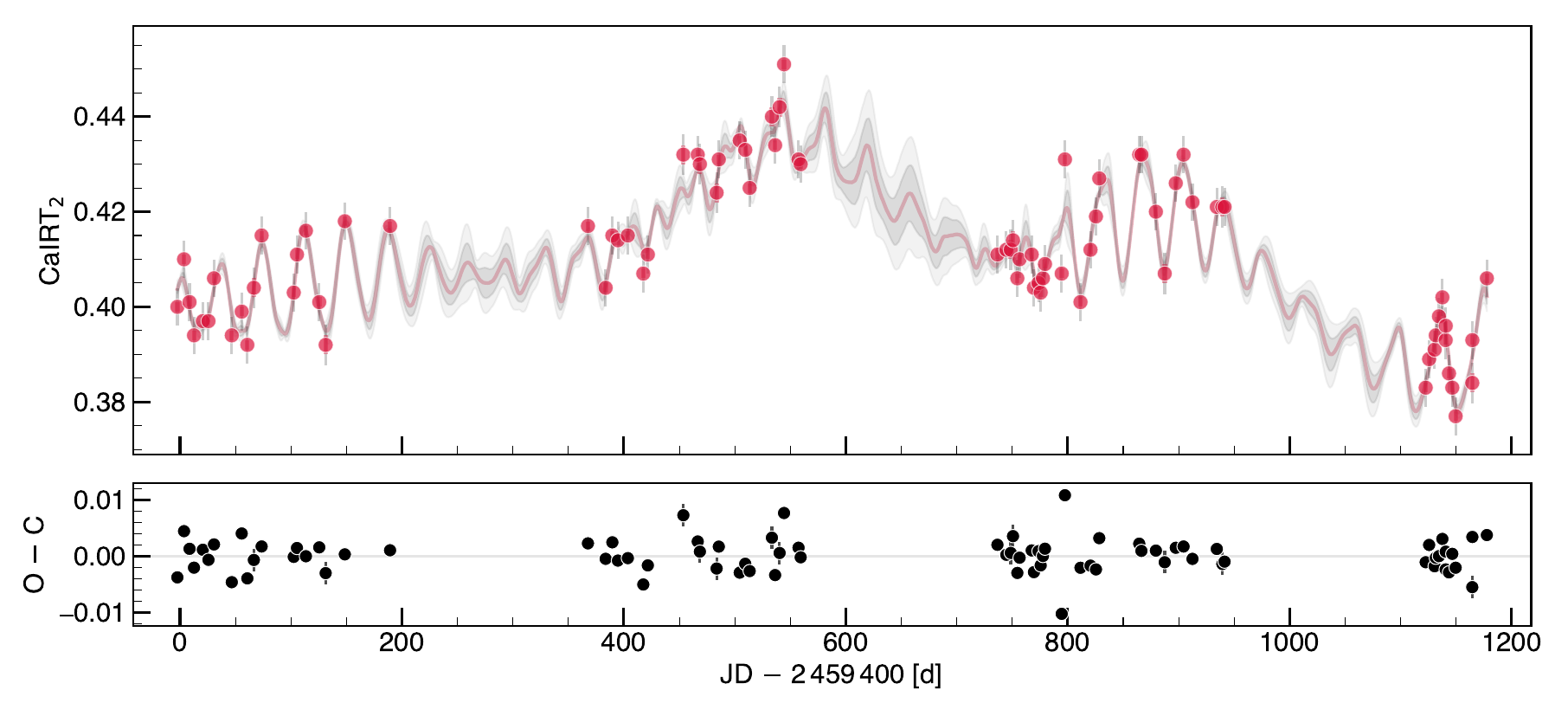}
\caption{\textit{Top:} CaIRT$_{\rm 2}$ measurements (colored dots) with the GP model (red line) and the 68.7\% and 95\% confidence intervals (dark and light grey regions). \textit{Bottom:} Residuals of the GP model. See Sect. 4.1.2.}
\label{fig:E8}
\end{figure*}

\begin{figure*}
\centering\includegraphics[width=1.0\textwidth{}]{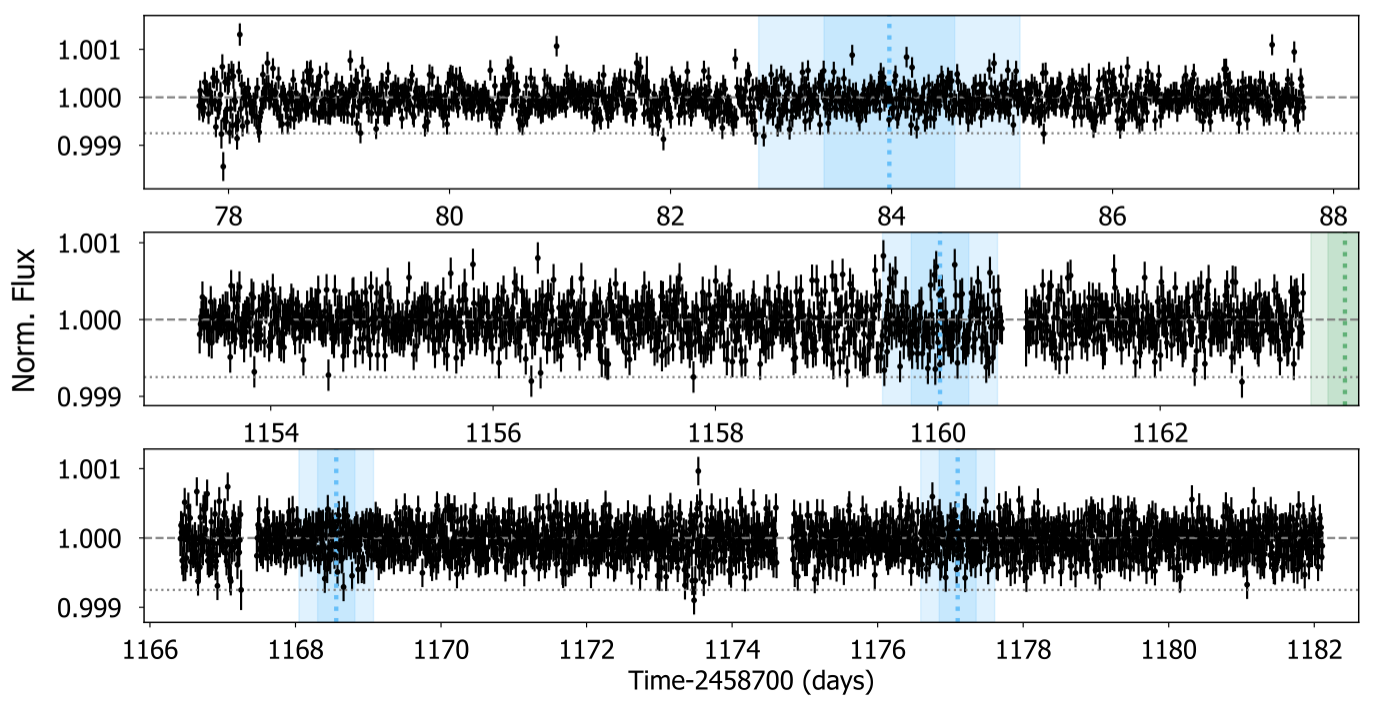}
\caption{TESS light curves for \host{}, including the second half of sector 17 (\textit{upper} panel) and sector 57 (two \textit{lower} panels). Each panel shows the different time series for each sector before and after the down-link gap. See caption from Fig. 4 (Sect. 4.2) for further details.}
\label{fig:E9}
\end{figure*}

\section{Additional tables}
\label{sec:addtables}

\setlength{\tabcolsep}{20pt}
\begin{table*}
\centering
\caption{Spectroscopic indicators from the CARMENES time series.} \label{tab:RVindicators}
\label{tab:outliers}
\begin{tabular}{ccccccc}
\hline
\hline \noalign{\smallskip}
BJD - 2\,400\,000 & FWHM & $\sigma_{\rm FWHM}$ 
                       & BIS & $\sigma_{\rm BIS}$ 
                        & DLW & \ldots \\ 
$[$d$]$  & [m\,s$^{-1}$] & [m\,s$^{-1}$] 
            & [m\,s$^{-1}$] & [m\,s$^{-1}$] 
            & [m$^2$\,s$^{-2}$] & 
            \\ \hline \noalign{\smallskip}

59397.642675    &       7652    &       13      &       83.0    &       7.7     &       -27.4   &         \\ \noalign{\smallskip}
59403.647097    &       7630    &       13      &       71.0    &       7.9     &       -14.7   &       \\ \noalign{\smallskip} 
59408.616823    &       7666    &       13      &       92.0    &       7.4     &       -22.1   &       \\ \noalign{\smallskip}
59412.629282    &       7670    &       13      &       80.0    &       7.1     &       -27.8   &   \\ \noalign{\smallskip}
59420.600123    &       7665    &       13      &       91.0    &       7.5     &       -17.7   &   \\ \noalign{\smallskip} 
59425.662880    &       7639    &       13      &       77.0    &       7.2     &       -23.7 &     \\ \noalign{\smallskip}
59430.654213    &       7675    &       13      &       89.0    &       5.7     &       -17.8   &   \\ \noalign{\smallskip} 
59446.684781    &       7655    &       13      &       71.0    &       6.3     &       -34.9   &       \\ \noalign{\smallskip}
59455.644156    &       7685    &       13      &       73.0    &       5.8     &       -12.2   &   \\ \noalign{\smallskip} 
59460.694884    &       7637    &       13      &       78.0    &       6.1     &       -29.3   &   \\ \noalign{\smallskip}
\ldots &                &               &               &               &               &         \\
\noalign{\smallskip}
\hline 
\end{tabular}
\tablefoot{Only the first ten rows and six columns are shown (the missing indicators are CRX, H$\alpha$, CaIRT1, CaIRT2, NaD1 and NaD2). The full version of this table is available at the CDS.}
\end{table*}

\setlength{\tabcolsep}{20pt}
\begin{table*}
\centering
\caption{CARMENES RV measurements extracted with \texttt{sBART} and the correspondingly applied nightly zero points and drifts. \label{tab:rv_sbart}} 
\begin{tabular}{ccccccc}
\hline
\hline \noalign{\smallskip}
BJD - 2\,400\,000 & RV & $\sigma_{\rm RV}$ & NZP & $\sigma_{\rm NZP}$ & drift & $\sigma_{\rm drift}$ \\
$[$d$]$  & [m\,s$^{-1}$] & [m\,s$^{-1}$] & [m\,s$^{-1}$] & [m\,s$^{-1}$] & [m\,s$^{-1}$] & [m\,s$^{-1}$]\\ \noalign{\smallskip}
\hline \noalign{\smallskip}
59397.642675    &       -23293.50       &       2.22    &       -0.07   &       0.53    &       1.22    &       0.35    \\ \noalign{\smallskip}
59403.647097    &       -23294.86       &       2.33    &       0.81    &       0.47    &       -2.14   &       0.57    \\ \noalign{\smallskip}
59408.616823    &       -23296.37       &       2.09    &       3.17    &       0.64    &       4.40    &       0.46    \\ \noalign{\smallskip}
59412.629282    &       -23293.75       &       2.14    &       2.94    &       0.52    &       -4.84   &       0.74    \\ \noalign{\smallskip}
59420.600123    &       -23296.27       &       2.14    &       5.08    &       0.62    &       1.13    &       0.61    \\ \noalign{\smallskip}
59425.662880    &       -23305.60       &       2.16    &       6.86    &       0.55    &       4.57    &       0.44    \\ \noalign{\smallskip}
59430.654213    &       -23291.87       &       1.84    &       3.45    &       0.53    &       4.41    &       0.41    \\ \noalign{\smallskip}
59446.684781    &       -23289.34       &       1.96    &       -0.18   &       0.62    &       17.42   &       0.65    \\ \noalign{\smallskip}
59455.644156    &       -23297.05       &       1.85    &       0.37    &       0.49    &       -15.03  &       0.78    \\ \noalign{\smallskip}
59460.694884    &       -23299.13       &       2.09    &       0.51    &       0.34    &       -6.29   &       1.08    \\ \noalign{\smallskip}
\ldots  & & & & & & \\
\noalign{\smallskip}
\hline 
\end{tabular}
\tablefoot{The full version of this table is available at the CDS.}
\end{table*}

\setlength{\tabcolsep}{20pt}
\begin{table}
\centering
\caption[]{Observing times discarded from the CARMENES time series. Further details in Sect. 2.1.}
\label{tab:F3}
\begin{tabular}{lc}
\hline
\hline \noalign{\smallskip}
BJD - 2\,400\,000 & Reason \\ \hline \noalign{\smallskip}
59483.611306 & Drift \\ \noalign{\smallskip}
59490.496133 & Drift \\ \noalign{\smallskip}
59568.353089 & Moon \\ \noalign{\smallskip}
59775.658997 & Drift \\ \noalign{\smallskip}
59809.637907 & S/N < 20 \\ \noalign{\smallskip}
59864.638976 & Moon \\ \noalign{\smallskip}
59896.425823 & Moon \\ \noalign{\smallskip}
59948.335626 & Moon \\ \noalign{\smallskip}
59951.349542 & Drift \\ \noalign{\smallskip}
60217.526971 & Moon \\ \noalign{\smallskip}
60270.343849 & S/N < 20 \\ \noalign{\smallskip}
60272.336330 & Moon \\ \noalign{\smallskip}
60274.457653 & Moon \\ \noalign{\smallskip}
60276.326409 & Moon \\ \noalign{\smallskip}
60302.379642 & Moon \\ \noalign{\smallskip}
60307.375308 & NZP \\ \noalign{\smallskip}
60555.574195 & NZP \\ \noalign{\smallskip}
\hline 
\end{tabular}
\end{table}

\begin{table*}
\centering
\caption{Summary of the ASAS-SN observations of \host{}. See Sect. 2.3.}
\label{tab:F4}
\setlength{\tabcolsep}{6pt}
\begin{tabular}{lllllll}
\hline \hline \noalign{\smallskip}
Start date  & End date    & Camera & Band & Station           & Location                                & $\#$Obs \\ \hline \noalign{\smallskip}
05 May 2013 & 29 Nov 2018 & bb     & $V$    & Brutus            & Haleakala Observatory                   & 421     \\ \noalign{\smallskip}
03 Dec 2018 & 30 Jan 2022 & bB     & $g$    & Brutus            & Haleakala Observatory                   & 161     \\ \noalign{\smallskip}
29 Oct 2017 & 11 Sep 2018 & bf     & $V$    & Cassius           & Cerro Tololo International Observatory  & 6       \\ \noalign{\smallskip}
02 Nov 2018 & 14 Aug 2024 & bF     & $g$    & Cassius           & Cerro Tololo International Observatory  & 68      \\ \noalign{\smallskip}
29 Nov 2017 & 16 Nov 2023 & bj     & $g$    & Henrietta Leavitt & McDonald Observatory                    & 126     \\ \noalign{\smallskip}
28 Nov 2017 & 03 Aug 2024 & bn     & $g$    & Cecilia Payne     & South African Astrophysical Observatory & 116     \\ \noalign{\smallskip}
22 Sep 2017 & 18 Aug 2024 & br     & $g$    & Bohdan Paczyński  & Cerro Tololo International Observatory  & 212     \\  \noalign{\smallskip} \hline
\end{tabular}
\end{table*}

\setlength{\tabcolsep}{12pt}
\begin{table}[]
    \centering
    \caption[]{General and dynamical properties of \host.}\label{tab:stellar_param}
    \begin{tabular}{@{}llc@{}}
    \hline \hline     \noalign{\smallskip}
    Parameter & Value & Ref. \\     \noalign{\smallskip} \hline 
    \noalign{\smallskip}
    \multicolumn{3}{c}{\em Basic identifiers} \\
    \noalign{\smallskip}
    Hipparcos  & \object{HIP~5957} & 1 \\
    Gliese  & \object{Gl~55.2} & 2 \\
    {\em TESS}  & \object{TIC~16917838} & 3 \\
    \gaia{} DR3\tablefootmark{a}  & 294517800251711616 & 4 \\
    Sp. type & K7\,V & 0\tablefootmark{b} \\
    \noalign{\smallskip}
    \multicolumn{3}{c}{\em Astrometry and kinematics} \\
    \noalign{\smallskip}
    $\alpha$ (J2016.0) & 01:16:27.72 & 4 \\ 
    $\delta$ (J2016.0) & +25:19:51.7 & 4 \\
    $l$ [deg] & 130.091481 & 4 \\ 
    $b$ [deg] & --37.189786 & 4 \\      
    $\varpi$ [mas] & 41.876 $\pm$ 0.021 & 4 \\
    $d$ [pc] & 23.880 $\pm$ 0.012 & 4 \\
    $\mu_{\alpha}\cos{\delta}$ [mas\,yr$^{-1}$] & 429.196 $\pm$ 0.019 & 4 \\
    $\mu_{\delta}$ [mas\,yr$^{-1}$] & --102.681 $\pm$ 0.017 & 4 \\
    RUWE & 1.057 & 4 \\
    $\gamma$ [km\,s$^{-1}$] & --23.10 $\pm$ 0.16 & 4 \\
    $U$ [km\,s$^{-1}$] & --23.959 $\pm$ 0.084 & 0 \\
    $V$ [km\,s$^{-1}$] & --48.627 $\pm$ 0.099 & 0 \\
    $W$ [km\,s$^{-1}$] & +9.523 $\pm$ 0.097 & 0 \\
    Gal. population & Thin disk & 0 \\
    \noalign{\smallskip}
    \multicolumn{3}{c}{\em Photometry} \\
    \noalign{\smallskip}   
    $V$ [mag] & 10.068 & 5 \\
    $G$ [mag] & 9.4775  $\pm$ 0.0028 & 4 \\
    $G_{BP}$ -- $G_{RP}$ [mag] & 1.73 & 4 \\
    $J$ [mag] & 7.464 $\pm$ 0.023 & 6 \\
    $K_s$ [mag] & 6.656 $\pm$ 0.018 & 6 \\
    \noalign{\smallskip}
    \multicolumn{3}{c}{\em Fundamental parameters} \\
    \noalign{\smallskip}  
    $T_{\rm eff}$ [K] & 4135 $\pm$ 36 & 0  \\
    $\log{g}$ &  5.07 $\pm$ 0.27 & 0 \\
    $[$Fe/H$]$ [dex] & --0.01 $\pm$ 0.09 & 0  \\
    $L_{\star}$ [$L_{\odot}$] & 0.10098 $\pm$ 0.00012 & 0 \\
    $M_{\star}$ [$M_{\odot}$] & 0.629 $\pm$ 0.017 & 0 \\
    $R_{\star}$ [$R_{\odot}$] & 0.619 $\pm$ 0.011  & 0 \\
    HZ range\tablefootmark{c} [au] & 0.25-0.65 & 0 \\
    HZ range [d] & 59-241 & 0 \\
    \noalign{\smallskip}
    \multicolumn{3}{c}{\em Activity} \\
    \noalign{\smallskip} 
    $v \sin{i}$ [km\,s$^{-1}$] & $<$2\tablefootmark{d} & 0 \\
    $P_{\rm rot}$ [d] & 37.4 $\pm$ 6.0 & 0 \\
    pEW(Ca$_{\rm II}$ IRT$_1$) [\AA] & +0.865 $\pm$ 0.006 & 0 \\
    log\,$R'_{HK}$ & --4.896 $\pm$ 0.003 & 0 \\ \noalign{\smallskip}
    \hline
    \end{tabular}
    \tablefoot{
    \tablefoottext{a}{DR3 contents can be accessed at: \url{https://www.cosmos.esa.int/web/gaia/dr3}}
    \tablefoottext{b}{The spectral classification of \host\ is at the boundary between late K and early M type, without a definitive distinction in the literature \citep[e.g.][]{reid95,gray03}.
    Our work classifies the host star as a K7 subtype.}
    \tablefoottext{c}{Optimistic range as defined by \cite{kopparapu13}.}
    \tablefoottext{d}{\cite{martinez10} determined a value of $v \sin{i} \leq$ 3.7\,km\,s$^{-1}$ using FOCES, a spectrograph with a resolution lower than CARMENES.}
    }
    \tablebib{
    (0) This work; 
    (1) \citet{ESA97}; 
    (2) \citet{gliese69}; 
    (3) \citet{stassun19}; 
    (4) \citet{gaia22}; 
    (5) \citet{koen10}; 
    (6) \citet{cutri03}. 
    }
\end{table}

{\footnotesize
\begin{table}
\centering
\caption[]{Prior distributions for the \texttt{kima} RV analysis. Subscript $i$ identifies the planet.}
\label{tab:F6}
\begin{tabular}{lc}
\hline
\hline \noalign{\smallskip}
Parameter & Prior \\ \hline \noalign{\smallskip}
$m$ [m$\,$s$^{-1}\,$d$^{-1}$] & $\mathcal{G}$(0, 0.01) \\
$q$ [m$\,$s$^{-1}\,$d$^{-2}$] & $\mathcal{G}$(0, 10$^{-5}$) \\\noalign{\smallskip}
$\gamma$ [m$\,$s$^{-1}$] & $\mathcal{U}$(-23\,341, -23\,241) \\ \noalign{\smallskip}
$\sigma_{\rm jit}$ [m$\,$s$^{-1}$] & $\mathcal{LU}$(0.01, 40) \\ \noalign{\smallskip}
$\nu^{a}$ & $\mathcal{LU}$(2, 1000) \\ \noalign{\smallskip}
$N_p$ & $\mathcal{U}$(0, 3) \\ \noalign{\smallskip}
\hline \noalign{\smallskip}
P$_{\rm i}$ [d] & $\mathcal{LU}$(1.0, 1157.9) \\ \noalign{\smallskip}
$K_{\rm i}$ [m$\,$s$^{-1}$] & $\mathcal{U}$(0, 27)  \\ \noalign{\smallskip} 
$e_{\rm i}$ & $\mathcal{K}^c$(0.881, 2.878) \\ \noalign{\smallskip}
$\phi_{\rm i}^b$ & $\mathcal{U}$(0, 2$\pi$) \\ \noalign{\smallskip}
$\omega_{\rm i}$ & $\mathcal{U}$(0, 2$\pi$) \\ \noalign{\smallskip}
\hline 
\end{tabular}
\tablefoot{
\tablefoottext{a}{Prior for the Student-t likelihood degrees of freedom.}
\tablefoottext{b}{True anomaly at the first observation.}
\tablefoottext{c}{$\mathcal{K}$ represents the Kumaraswamy prior (\citealt{kumapr}).}
}
\end{table}
}

\setlength{\tabcolsep}{18pt}
\begin{table*}[]
\centering
\caption{Prior and posterior distributions for the best RV model (2p1c2c), with inferred planetary properties.}
\label{tab:F7}
\begin{tabular}{lll}
\hline
\hline \noalign{\smallskip}
Parameter & Prior$^a$ & Posterior \\ \hline \noalign{\smallskip}
$m$ [m$\,$s$^{-1}\,$d$^{-1}$] & $\mathcal{U}$(-0.05, 0.05) & $-($7.2\,$\pm$\,3.2)\,10$^{-3}$ \\ 
$q$ [m$\,$s$^{-1}\,$d$^{-2}$] & $\mathcal{U}$(-0.01, 0.01) & $-($7.6\,$\pm$\,2.6)\,10$^{-6}$ \\ 
\noalign{\smallskip}
$\gamma$ [m$\,$s$^{-1}$] & $\mathcal{U}$(--23\,330.6, --23\,255.7) & --23\,295.14$^{+0.85}_{-0.84}$ \\ \noalign{\smallskip} 
$\sigma_{\rm jit}$ [m$\,$s$^{-1}$] & $\mathcal{MLU}$(1.9, 24.9) & 2.17$_{-0.32}^{+0.34}$ \\ \noalign{\smallskip} 
\hline \noalign{\smallskip}
$P_{\rm b}$ [d] & $\mathcal{LU}$(1.1, 100.0) & 8.5399$_{-0.0036}^{+0.0037}$ \\ \noalign{\smallskip} 
$K_{\rm b}$ [m$\,$s$^{-1}$] & $\mathcal{U}$(0.0, 14.0) & 3.75\,$\pm$\,0.48 \\ \noalign{\smallskip}
$T_{\rm 0,b}$ [d] & $\mathcal{U}$(2\,459\,397.6, 2\,459\,497.6) & 2\,459\,484.26$_{-0.30}^{+0.29}$ \\ \noalign{\smallskip} 
$e_{\rm b}$ & Fixed to 0 & \\ \noalign{\smallskip}
\hline \noalign{\smallskip}
$P_{\rm c}$ [d] & $\mathcal{LU}$(1.1, 100.0) & 29.671$_{-0.052}^{+0.050}$ \\ \noalign{\smallskip}
$K_{\rm c}$ [m$\,$s$^{-1}$] & $\mathcal{U}$(0.0, 14.0) & 3.50\,$\pm$\,0.45 \\ \noalign{\smallskip}
$T_{\rm 0,c}$ [d] & $\mathcal{U}$(2\,459\,397.6, 2\,459\,497.6) & 2\,459\,448.2\,$\pm$\,1.2 \\ \noalign{\smallskip}
$e_{\rm c}$ & Fixed to 0 &  \\ \noalign{\smallskip}
\hline
\hline \noalign{\smallskip}
Planet & Parameter & Inferred value \\ \noalign{\smallskip}
\hline  \noalign{\smallskip}
\multirow{5.5}{*}{\planetb} & Orbit semi-major axis, $a_{\rm b}$ [au] & 0.07006\,$\pm$\,0.00043 \\ \noalign{\smallskip}
& Relative orbital separation, $a_{\rm b}/R_{\star}$ & 24.33$^{+0.33}_{-0.32}$ \\ \noalign{\smallskip}
& Minimum mass, $m_{\rm b}\sin{i_{\rm b}}$ [M$_{\oplus}$] & 8.80\,$\pm$\,0.76 \\ \noalign{\smallskip}
& Isolation flux, $S_{\rm b}$ [$S_{\oplus}$] & 20.56\,$\pm$\,0.25 \\ \noalign{\smallskip}
& Equilibrium temperature, $T_{\rm eq,\,b}$ [K] & 594.4\,$\pm$\,5.3 \\ \noalign{\smallskip}
\hline \noalign{\smallskip}
\multirow{5.5}{*}{\planetc} & Orbit semi-major axis, $a_{\rm c}$ [au] & 0.16071\,$\pm$\,0.00099 \\ \noalign{\smallskip}
& Relative orbital separation, $a_{\rm c}/R_{\star}$ & 55.81$^{+0.75}_{-0.73}$ \\ \noalign{\smallskip}
& Minimum mass, $m_{\rm c}\sin{i_{\rm c}}$ [M$_{\oplus}$]& 12.4\,$\pm$\,1.1 \\ \noalign{\smallskip}
& Isolation flux, $S_{\rm c}$ [S$_{\oplus}$] & 3.910\,$\pm$\,0.047 \\ \noalign{\smallskip}
& Equilibrium temperature, $T_{\rm eq,\,c}$ [K] & 392.5\,$\pm$\,3.5 \\ \noalign{\smallskip}
\hline
\end{tabular}
\tablefoot{
\tablefoottext{a}{See Sect.~\ref{sec:emcee} for details on the selection of these priors.}
} 
\end{table*}

\setlength{\tabcolsep}{20pt}
\begin{table*}
\centering
\caption[]{Prior and posterior distributions for the 2p1c2c + GP model for the \sbart\ RVs. See Sect. 4.1.2.}
\label{tab:F8}
\begin{tabular}{lcc}
\hline
\hline \noalign{\smallskip}
Parameter\,$^a$ & Prior & Posterior \\ \hline \noalign{\smallskip}
$m$ [m$\,$s$^{-1}\,$d$^{-1}$] & $\mathcal{U}$(-0.05, 0.05) & $-($7.6\,$\pm$\,4.3)\,10$^{-3}$ \\ \noalign{\smallskip}
$q$ [m$\,$s$^{-1}\,$d$^{-2}$] & $\mathcal{U}$(-0.01, 0.01) & $-($7.9\,$\pm$\,3.6)\,10$^{-6}$ \\ 
\noalign{\smallskip}
$\gamma$ [m$\,$s$^{-1}$] & $\mathcal{U}$(--23\,330.6, --23\,255.7) & --23\,295.3$^{+1.1}_{-1.2}$\\ \noalign{\smallskip}
$\sigma_{\rm jit}$ [m$\,$s$^{-1}$] & $\mathcal{MLU}$(1.9, 24.9) & 2.11$^{+0.37}_{-0.35}$ \\ \noalign{\smallskip} 
\hline \noalign{\smallskip}
$m_{PX}$ & $\mathcal{U}$(-0.05,\,0.05) & (8.9~$\pm$~4.0)\,10$^{-5}$\\ \noalign{\smallskip}
$q_{PX}$ & $\mathcal{U}$(-0.01,\,0.01) & --8.0~$\pm$~3.2)\,10$^{-8}$ \\ \noalign{\smallskip}
$\gamma_{\rm{PX}}$ & $\mathcal{U}$(0.30, 0.53) & 0.3978$^{+0.0100}_{-0.0098}$ \\ \noalign{\smallskip}
$\sigma_{\rm jit,\,PX}$ & $\mathcal{MLU}$(10$^{-3}$, 7.4\,10$^{-2}$) & (3.86$^{+0.70}_{-0.68})10^{-3}$ \\ \noalign{\smallskip}
\hline \noalign{\smallskip}
$P_{\rm b}$ [d] & $\mathcal{LU}$(1.1, 100.0) & 8.5400$^{+0.0037}_{-0.0036}$ \\ \noalign{\smallskip}
$K_{\rm b}$ [m$\,$s$^{-1}$] & $\mathcal{U}$(0.0, 14.0) & 3.78~$\pm$~0.50 \\ \noalign{\smallskip}
$T_{\rm 0,b}$ [d] & $\mathcal{U}$(2\,459\,397.6, 2\,459\,497.6) & 2\,459\,415.94$_{-0.33}^{+0.31}$\\ \noalign{\smallskip} 
$e_{\rm b}$ & Fixed to 0 &  \\ \noalign{\smallskip}
\hline \noalign{\smallskip}
$P_{\rm c}$ [d] & $\mathcal{LU}$(1.1, 100.0) & 29.675$^{+0.052}_{-0.054}$ \\ \noalign{\smallskip}
$K_{\rm c}$ [m$\,$s$^{-1}$] & $\mathcal{U}$(0.0, 14.0) & 3.49~$\pm$~0.48 \\ \noalign{\smallskip}
$T_{\rm 0,c}$ [d] & $\mathcal{U}$(2\,459\,397.6, 2\,459\,497.6) & 2\,459\,448.2~$\pm$~1.2 \\ \noalign{\smallskip}
$e_{\rm c}$ & Fixed to 0 &  \\ \noalign{\smallskip}
\noalign{\smallskip}
\hline \noalign{\smallskip}
$\eta_{\rm1,\,RV}$ [m$\,$s$^{-1}$] & $\mathcal{U}$(0.0, 14.0) & 0.70$_{-0.50}^{+0.80}\,^b$ \\ \noalign{\smallskip}
$\eta_{\rm1,\,PX}$ & $\mathcal{U}$(0.0, 0.06) & (1.26$^{-0.24}_{+0.43}) 10^{-2}$ \\ \noalign{\smallskip}
$\eta_{\rm2}$ [d] & $\mathcal{U}$(56.1, 3540.2) & 83$^{-17}_{+23}$ \\ \noalign{\smallskip}
$\eta_{\rm3}$ [d] & $\mathcal{G}_t\left(37.4, 6.0\right)$ & 37.90$^{-0.99}_{+0.86}$ \\ \noalign{\smallskip}
$\eta_{\rm4}$ & $\mathcal{LU}\left(10^{-2},~10.0\right)$ & 0.75$^{-0.21}_{+0.37}$ \\ \noalign{\smallskip}
\hline 
\end{tabular}
\tablefoot{
\tablefoottext{a}{PX denotes the proxy used, in this case being CaIRT$_2$.}
\tablefoottext{b}{Mode of the marginalized posterior distribution at 0.
}
}
\end{table*}

\begin{table*}
\centering
\caption[]{Integration times for the \texttt{LIFEsim} simulations. See Sect. 5.4.2.}
\label{tab:F9}
\begin{tabular}{llll}
\hline
\hline \noalign{\smallskip}
Planet type & Configuration & t$_{\rm int,\,\planetb}$  &  t$_{\rm int,\,\planetc}$  \\ \hline \noalign{\smallskip}
\multirow{0}{*}{Super-Earth} & Optimistic & 15 min & 45 min \\ \noalign{\smallskip}
& Baseline & 1 h & 4 h \\ \noalign{\smallskip}
& Pessimistic & 13 h & 44 h \\ \noalign{\smallskip}
\noalign{\smallskip}
\noalign{\smallskip}
\multirow{0}{*}{Mini-Neptune} & Optimistic & 5 min & 10 min \\ \noalign{\smallskip}
& Baseline & 20 min & 50 min \\ \noalign{\smallskip}
& Pessimistic & 4 h & 9 h \\ \noalign{\smallskip}
\hline 
\end{tabular}
\end{table*}

\setlength{\tabcolsep}{20pt}
\begin{table*}
\centering
\caption[]{Priors and posteriors of the analysis of the TESS light curve assuming Gaussian priors on the period from the RV analysis. See Sect. 4.2.}
\label{tab:F10}
\begin{tabular}{lcc}
\hline 
\hline \noalign{\smallskip} 
Parameter & Prior & Posterior \\
\hline 
\textit{Inferred parameters} & & \\
Orbital period, $P_{\rm b}$ [d] & $\mathcal{G}$(29.64, 0.040) & $29.640^{+0.039}_{-0.040}$ \\
Time of mid-transit, $T_{\rm 0,b}-2\,400\,000$ [d] & $\mathcal{U}$(58\,764.0, 58\,773.0) & $58\,768.6582^{+3.0205}_{-0.0047}$ \\
Orbital inclination, $i_{\rm b}$ [deg.] & $\mathcal{U}$(70.0, 90.0) & $89.210^{+0.139}_{-0.075}$ \\
Planet radius, $R_b$ [$R_{\oplus}$] & $\mathcal{U}$(0.0, 10.0) & $1.69\pm0.12$ \\
Stellar radius, $R_{\star}$ [$R_{\odot}$] & $\mathcal{G}$(0.619, 0.011) & $0.619^{+0.011}_{-0.011}$ \\
Stellar mass, $M_{\star}$ [$M_{\odot}$] & $\mathcal{G}$(0.629, 0.017) & $0.631^{+0.018}_{-0.016}$ \\ \noalign{\smallskip} \hline 
\textit{Derived parameters} & & \\
Transit depth, $\Delta_{b}$ [ppt] & (derived) & $0.60^{+0.13}_{-0.21}$ \\
Orbit semi-major axis, $a_{b}$ [au] & (derived) & $0.1608^{+0.0014}_{-0.0016}$ \\
Relative orbital separation, $a_{b}/R_{\star}$ & (derived) & $55.9^{+1.1}_{-1.1}$ \\
Impact parameter, $b_{b}/R_{\star}$ & (derived) & $0.769^{+0.136}_{-0.065}$ \\
Incident Flux, $F_{\rm inc,b}$ [$F_{{\rm inc},\oplus}$] & (derived) & $3.89^{+0.43}_{-0.39}$ \\
Equilibrium temperature, $T_{\rm eq,b}$ [K] & (derived) & $356.7^{+9.4}_{-9.3}$ \\ \noalign{\smallskip}
\hline 
\end{tabular}
\end{table*}

\end{document}